# Non-associated Flow Rule-Based Elasto-Viscoplastic Model for Clay

**Mohammad N. Islam * and Carthigesu T. Gnanendran**

School of Engineering and Information Technology, University of New South Wales, Canberra, ACT 2612, Australia; r.gnanendran@adfa.edu.au

* Correspondence: Mohammad.Islam@netl.doe.gov

**Abstract:** We develop a non-associated flow rule (NAFR)-based elasto-viscoplastic (EVP) model for isotropic clays. For the model formulation, we introduce the critical state soil mechanics theory (CSSMT), the bounding surface theory and Perzyna's overstress theory. The NAFR based EVP model comprises three surfaces: the potential surface, the reference surface and the loading surface. Additionally, in the model formulation, assuming the potential surface and the reference surface are identical, we obtain the associated flow rule-based EVP model. Both EVP models require seven parameters and five of them are identical to the Modified Cam Clay model. The other two parameters are the surface shape parameter and the secondary compression index. Moreover, we introduce the shape parameter in the model formulation to control the surface shape and to account for the overconsolidation state of clay. Additionally, we incorporate the secondary compression index to introduce the viscosity of clay. Also, we validate the EVP model performances for the Shanghai clay, the San Francisco Bay Mud (SFBM) clay and the Kaolin clay. Furthermore, we use the EVP models to predict the long-term field monitoring measurement of the Nerang Broadbeach roadway embankment in Australia. From the comparison of model predictions, we find that the non-associated flow rule EVP model captures well a wide range of experimental results and field monitoring embankment data. Furthermore, we also observe that the natural clay exhibits the flow rule effect more compared to the reconstituted clay.

## 1. Introduction

In a saturated clay medium, the liquid phase occupies the interparticle void spaces of the solid phase. When such a clay deposit experiences external loading, saturated soil either exhibits slow loading conditions or fast loading conditions. The first one is also known as the drained behavior, while the latter one represents undrained behavior. In most cases, the liquid removal of soft clays is not instantaneous due to its low permeability, compressibility characteristics and the viscous nature [1,2]. Therefore, depending on the loading state and the in situ condition, in many cases, the deformation of clay may continue for a long time [3]. For example, in August 1173, the Leaning Tower of Pisa in Italy was constructed on a highly compressible clay deposit, and it displaced horizontally to the magnitude of 4.7 m in 1990, which is also increasing 1.5 mm/year [4]. A similar situation also may happen in any geotechnical structure founded on soft clay (see also Brand and Brenner [2]). In this regard, the viscosity of clay most often contributes to the long-term time-dependent creep of clay, and subsequent damage to the structure, which requires billions of dollars in annual maintenance costs [5].

Additionally, only in the USA, the sum of onshore abandoned wells is about 3 million, and every year approximately 40,000 new deep boreholes are drilled [6]. The estimated cost using the bentonite clay-based plug of those abandoned wells is about $160,000 per shallow borehole, which also

demonstrates a billion-dollar legacy (see also Islam et al. [6]). Furthermore, in an engineered barrier system for hazardous wastes (e.g., nuclear waste), the application of bentonite clay is also common and the clay type barrier requires 100's of years of monitoring due to the sensitive nature of the deposited materials [7]. Additionally, it is important to incorporate time-dependent viscous responses of clay in a constitutive model formulation to obtain realistic hydro-mechanical behavior [8]. Therefore, clay-based research is the active field of interest and the motivation of the present paper.

From the early 19th century to the present time, to illustrate the clay behavior, a myriad of coupled constitutive models have been developed, including viscous-inclusive (e.g., Adachi and Okano [9]) and viscous-exclusive (e.g., Roscoe and Burland [10]) models. However, we limited our discussion only to the first group of models. In this regard, among others, Liingaard et al. [11] and Chaboche [12] provided literature reviews. Nevertheless, to avoid the mathematical formulation complexity, in most cases, the time-dependent constitutive models are limited to the associated flow rule (AFR). But, capturing the legitimate behavior of soft clay, the non-associated flow rule (NAFR) is imperative (Zienkiewicz et al. [13]). In the literature, there are a couple of NAFR-based EVP models, where the number of material parameters ranged between six to 44. In this regard, Islam et al. [14] also presented a summary of NAFR-based elasto-viscoplastic (EVP) models. It is worth mentioning that a constitutive model with too many model parameters may capture geomaterials' behavior very well, but most often, their practical applications are not convenient [15]. Additionally, for the engineering application of any EVP model, the model formulation simplicity, its finite element implementation and the objective determination of model parameters are essential.

In this paper, we develop a non-associated flow rule-based elasto-viscoplastic (EVP) model considering the Modified Cam Clay (MCC) model [10] framework, Perzyna's overstressed theory [16], the Borja and Kavazanjian [17] concept, the bounding surface theory and the mapping rule (see also Hashiguchi [18]). The EVP model here requires a total of seven parameters, and among them, five parameters are identical to the MCC model. The other parameters are the secondary compression index and the surface shape parameter. We also introduce a non-linear secondary compression index to account for the viscosity of clay. Additionally, the shape of surfaces in this paper is different than the original MCC model or EVP model with the MCC model equivalent surface (see also Islam et al. [14]). We consider a non-circular surface in the π-plane. Also, we introduce a composite boundary surface, and the shape parameter to control the bounding surfaces' shape.

Furthermore, we also discuss the importance of the non-associated flow rule, the non-circular shape surface and the composite bounding surface. For validation of the EVP model, we compare numerical results with a wide variety of laboratory observed test data considering the Shanghai clay, the San Francisco Bay Mud clay and the Kaolin Clay. Moreover, after validation of the developed non-associated flow rule EVP model, we also implement it in a coupled finite element solver named *A F*inite *E*lement *N*umerical *A*lgorithm (AFENA) [19]. Additionally, for a field application of the developed EVP model, we compare the predicted response with the long-term monitoring measured response of the Nerang Broadbeach Roadway (NBR) embankment in Australia [20].

## 2. Importance of the Non-Associated Flow Rule

In this paper, we formulate the non-associated flow rule elasto-viscoplastic model considering the Modified Cam Clay (MCC) model framework, which was formulated considering the associated flow rule. Therefore, at first, we discuss the importance of the non-associated flow rule in the context of the MCC model as follows.

In the triaxial space, the incremental strain ($d\varepsilon$) is divided into the volumetric ($d\varepsilon_v$) and the deviatoric ($d\varepsilon_q$) component, while each of them is also comprised of the elastic part and the inelastic part (e.g., plastic or viscoplastic) as follows (see also Yu [15]; Roscoe and Burland [10]):

$$d\varepsilon = d\varepsilon_v + d\varepsilon_q, \tag{1}$$

$$d\varepsilon_v = d\varepsilon_v^e + d\varepsilon_v^p = d\varepsilon_1 + 2d\varepsilon_3, \tag{2}$$

$$d\varepsilon_q = d\varepsilon_q^e + d\varepsilon_q^p = \frac{2}{3}(d\varepsilon_1 - d\varepsilon_3), \tag{3}$$

where, $d\varepsilon_v$ and $d\varepsilon_q$ are the total volumetric strain and the total deviatoric strain, respectively; while superscript $e$ and $p$ represent their corresponding elastic and plastic components. $d\varepsilon_1$ and $d\varepsilon_3$ are the incremental major strain and the incremental minor principal strain, respectively. Again, for the 1D consolidation by noting $\varepsilon_3 = 0$ in Equation (3), the ratio of the incremental volumetric strain to the incremental deviatoric strain is obtained as:

$$\frac{d\varepsilon_v}{d\varepsilon_q} = \frac{d\varepsilon_v^p + d\varepsilon_v^e}{d\varepsilon_q^p + d\varepsilon_q^e} = \frac{3}{2}. \tag{4}$$

Additionally, by neglecting the relative magnitude of the elastic shear strain to the plastic shear strain, Yu [15] presented Equation (4) as:

$$\frac{d\varepsilon_v^p}{d\varepsilon_q^p} = \frac{3}{2}\frac{\lambda-\kappa}{\lambda}. \tag{5}$$

Also, Yu [15] presented the stress dilatancy relation as follows:

$$\frac{d\varepsilon_v^p}{d\varepsilon_q^p} = \frac{M^n - \acute{\eta}^n}{m\acute{\eta}^{n-1}}. \tag{6}$$

In Equation (6), $m, n$ are the material constants. $M$ and $\acute{\eta}$ are the critical state line (CSL) slope, and the stress ratio, respectively (see also Yu [15]). Furthermore, for the MCC model, $m = n = 2$. In Equation (5), $\frac{\lambda-\kappa}{\lambda}$ is a constant and assuming $\frac{\lambda-\kappa}{\lambda} \approx 0.80$ (see Schofield and Wroth [21]), the magnitude of the incremental volumetric plastic strain rate ratio $\left(\frac{d\varepsilon_v^p}{d\varepsilon_q^p}\right)$ becomes 1.2. By substituting $\frac{d\varepsilon_v^p}{d\varepsilon_q^p} = 1.2$ in Equation (6), the stress ratio for the normal compression ($\acute{\eta}_{0,nc}$) and the $K_0$ conditions become 0.40. However, $\acute{\eta}_{0,nc} = \frac{3M}{6-M} = 0.60$ (see also Yu [15]). Additionally, McDowell and his co-workers also proposed that $\acute{\eta}_{0,nc} = 0.6M = 0.60$ (see McDowell and Hau [22]). Therefore, it implies that in the associated flow rule condition, the predicted $\acute{\eta}_{0,nc}$ is too low.

Moreover, the yield surface of the MCC model is given by:

$$\left(1 + \left(\frac{q}{Mp}\right)^2\right)p = p_c, \tag{7}$$

where, $dp, dq$ and $dp_c$ are the differential quantities of $p, q$ and $p_c$ respectively. It is to note that during the undrained triaxial tests, the magnitude of $dp$ is a negative quantity for the normally consolidated clay. Therefore, in the 'wet' part when $\acute{\eta} < M$ (see also Schofield and Wroth [21]):

$$\left(\left(\frac{\eta}{M}\right)^2 - 1\right)dp > 0. \tag{8}$$

As a result, if the yield surface is not permitted to contract, as $dp_c > 0$ and $dq > 0$, the strain-softening phenomena will not occur.

In addition, when the stress state reaches the yield condition, a material is subjected to the plastic deformation, which is known as the plastic flow (see also Hashiguchi [18]). The plastic potential function illustrates the post-yield and the failure behavior of soils, and the plastic strain increment is normal to it. If the plastic strain is calculated on any surface other than the potential surface as in the associated flow rule condition, the predicted plastic shear strain is too high (see also Yu [15]; McDowell and Hau [22]).

It is worth mentioning that in the associated flow rule condition, the adopted yield surface over-estimates failure stresses on the dry side, and the bifurcation is not possible in the hardening regime (see also Yu [15]; Schofield and Wroth [21]). Additionally, in the associated flow rule (AFR), the yield surface and the potential surface are the same. From the evidence of the triaxial test results in the literature, it is observed that in the AFR condition, if the contraction of the yield surface with hardening is not allowed, the deviatoric strain is suppressed. Therefore, strain hardening behavior in the drained triaxial shearing entailed that the yield surface shrinkage with the hardening should not be permitted. Hence, in the associated flow rule condition, the strain hardening in the drained triaxial test and the strain softening in the undrained tests will not transpire. For this reason, to explain the plastic volume expansion and the softening behavior of plastically compressible materials, e.g., soft clay, a plastic potential surface rather than the yield surface is essential (see also Yu [15]).

To resolve the limitations mentioned above in the Modified Cam Clay (MCC) framework, we introduce the non-associated flow rule-based elasto-viscoplastic (EVP) model. Additionally, the MCC model is not capable of modeling the long-term viscous behavior of soft clay (see Islam and Gnanendran [23]), which is also a motivation of the EVP model formulation herein. Moreover, it is noteworthy that the MCC model yield surface (see also Equation (7)) assume the Von-Mises type circular surface in the π-plane. Thereby, except in the triaxial compression state, in any other loading state, the MCC model or equivalent EVP model with the circular type surface overestimate the stress (see also Islam and Gnanendran [23]; Islam et al. [14]). Furthermore, from the literature, comparing the single surface-based model with the composite surface-based model, it is observed that the first one underpredicts the soil state in the overconsolidated state compared to the latter one.

Moreover, from experimental evidence, we find two phenomena (see Schofield and Wroth [21]). The first one is for the overconsolidated clay, and it is evident that with increases in the overconsolidation ratio (OCR), the strength locus for the overconsolidated clay approach to the zero-tension line. The second case is for the normally consolidated clay where the strength locus intersects the critical state line in the mean pressure-deviatoric (*p-q*) plane. In this paper, to resolve such shortcomings, we also revise the MCC model's single yield surface, and we discuss details of them in the model formulation section.

### 3. Numerical Modeling

We assume that the porous media: (i) comprises of two phases (the solid phase and the liquid phase (see also Figure 1), (ii) fully saturated, (iii) obeys the small deformation, (iv) fluids follow Darcy's law, (v) supports the isotropic state, the static equilibrium and the isothermal equilibrium conditions, (vi) individual soil grains and the liquid phase are incompressible. Additionally, we ignore the geochemical effect, interaction forces and dynamic actions. Thereby, we also assume each phase density is constant. Furthermore, we consider the soil mechanics' basic principles (see Terzaghi [1]). First, deformations of solid-phase originate due to the liquid-phase removal and rearrangement of the solid's grain. Second, the total stress is the summation of the load carried by the soil (effective stress) and the fluid (pore pressure) (see also Schofield and Wroth [21]). Introducing Terzaghi's effective stress concept [1], we couple the solid phase and the liquid phase.

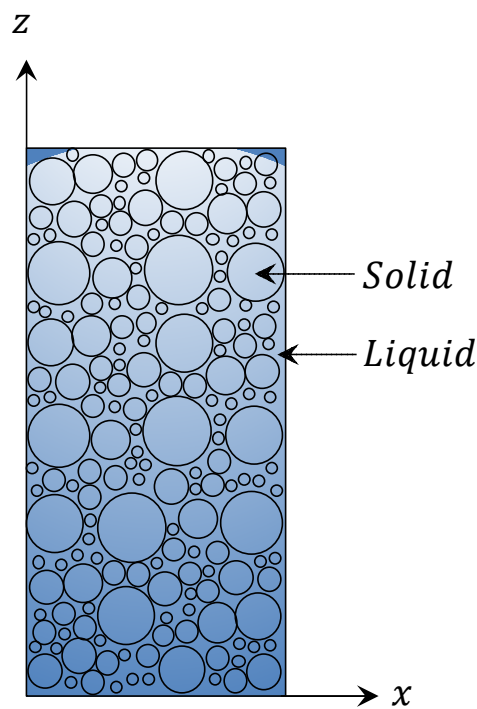

**Figure 1.** Two phases of representative elementary volume.

*3.1. Governing Equations*

Considering the assumptions above, we present governing equations as follows (see also Bear and Bachmat [24]).

Momentum balance equation:
$$div\boldsymbol{\sigma} + \rho \boldsymbol{g} = \boldsymbol{0}. \tag{9}$$

Mass balance equation:
$$\frac{\partial(\varphi\rho_l)}{\partial t} + div\,(\varphi\rho_l\boldsymbol{v}^l) = 0, \tag{10}$$
$$\frac{\partial((1-\varphi)\rho_s)}{\partial t} + div\,((1-\varphi)\rho_s\boldsymbol{v}^s) = 0, \tag{11}$$

where '$div$' is the divergence operator. $\boldsymbol{\sigma}$ is the total stress tensor. $\rho$, $\rho_l$ and $\rho_s$ are the total density of the porous medium, the liquid phase density and the solid phase density. $\boldsymbol{g}$ is the gravitational acceleration acting along the z-axis (see also Figure 1). $\varphi$ is the porosity of the porous medium. $\frac{\partial(\cdot)}{\partial t}$ is the time derivative. $\boldsymbol{v}^s$ and $\boldsymbol{v}^l$ are the solid phase and the liquid phase velocity, respectively. Additionally, $\rho\boldsymbol{g}$ is also known as the body force $\left(\boldsymbol{b} = [b_x, b_y, b_z]^T\right)$, where '$T$' represents transpose.

*3.2. Constitutive Assumptions*

In Equation (9), $\boldsymbol{\sigma}$ can be defined as (see Bear and Bachmat [24]):
$$\boldsymbol{\sigma} = \boldsymbol{\sigma}' + p_l\boldsymbol{I}. \tag{12}$$

Here, $\boldsymbol{I}$ is the identity tensor. $\boldsymbol{\sigma}'$ is the effective stress tensor and $p_l$ is the liquid pressure. Additionally, in Equations (10) and (11), we assume $\rho_l$ and $\rho_s$ are constant. Hence, after the summation of Equations (10) and (11), we obtain:
$$div\,[\varphi(\boldsymbol{v}^l - \boldsymbol{v}^s)] + div\,\boldsymbol{v}^s = 0. \tag{13}$$

For the two-phases porous media, Bear and Bachmat [24] presented the relation between $\boldsymbol{v}^l$ and $\boldsymbol{v}^s$ using the relative velocity ($\boldsymbol{w}_l$) term and the relative specific discharge ($\boldsymbol{q}_{rl}$) as follows:
$$\boldsymbol{v}^l = \boldsymbol{w}_l + \boldsymbol{v}^s, \tag{14}$$
$$\boldsymbol{w}_l = \frac{\boldsymbol{q}_{rl}}{\varphi} = \boldsymbol{v}^l - \boldsymbol{v}^s, \tag{15}$$
$$\boldsymbol{w}_l = -\frac{\boldsymbol{K}}{\gamma_l}(\nabla p^l - \rho_l \boldsymbol{g}), \tag{16}$$

where, $\nabla$ is the gradient operator. $\boldsymbol{K} = \frac{k\rho_l g}{\mu_l}$ is the hydraulic conductivity which depends on the fluid phase or fluidity $\left(\frac{\rho_l}{\mu_l}\right)$ and the specific permeability tensor of soil ($\boldsymbol{k}$). Additionally, $\rho_l \boldsymbol{g}$ is the liquid phase volumetric weight ($\gamma_l$), which also express as $\boldsymbol{b}_l = [0,0,\gamma_l]^T$, where '$T$' represents transpose. $\mu_l$ is the liquid phase viscosity and $\boldsymbol{g}$ is the gravitational acceleration.

Again, in Equation (13), '$div\,\boldsymbol{v}^s$' term is given by (see Bear and Bachmat [24]):
$$div\,\boldsymbol{v}^s = -\frac{\partial \varepsilon_v}{\partial t}. \tag{17}$$

Here, $\varepsilon_v$ is the volumetric strain and reads:
$$\dot{\varepsilon}_v = \frac{\partial(tr\varepsilon)}{\partial t}. \tag{18}$$

Assuming the small deformation, we also obtain the strain-displacement relation as follows:
$$\dot{\varepsilon}_{ij} = \frac{1}{2}\left(\frac{\partial \dot{u}_i}{\partial x_j} + \frac{\partial \dot{u}_j}{\partial x_i}\right), \tag{19}$$

where $\boldsymbol{u}$ is the displacement component.

Substituting, Equations (16) to (18) into Equation (13), then rearranging, we obtain:
$$div\left[\frac{\boldsymbol{K}}{\gamma_l}(\nabla p^l - \rho_l \boldsymbol{g})\right] + \frac{\partial \varepsilon_v}{\partial t} = 0. \tag{20}$$

It is worth mentioning that in the non-associated flow rule EVP model formulation, we assume that $\dot{\varepsilon}_{ij}$ (see Equations (17) to (19)) consists of the elastic component and the viscoplastic component as follows.

3.2.1. Strain Rate Tensor of the EVP Model

The total strain rate $(\dot{\varepsilon}_{ij})$ in the non-associated flow rule-based EVP model which is given by (see also Lubliner [25]):
$$\dot{\varepsilon}_{ij} = \dot{\varepsilon}_{ij}^e + \dot{\varepsilon}_{ij}^{vp}, \tag{21}$$

where, $\dot{\varepsilon}_{ij}^e$ and $\dot{\varepsilon}_{ij}^{vp}$ are the elastic strain rate tensor and the viscoplastic strain rate tensor, respectively. We obtain $\dot{\varepsilon}_{ij}^e$ as follows (see also Lubliner [25]):

$$\dot{\varepsilon}_{ij}^e = S_{ijkl}\dot{\sigma}'_{kl}, \tag{22}$$

where, $\dot{\sigma}'_{kl}$ is the effective stress tensor. $S_{ijkl}$ is the fourth-order compliance tensor and written as:

$$S_{ijkl} = \frac{1+\nu}{2E}\left(\frac{-2\nu}{1+\nu}\delta_{ij}\delta_{kl} + \delta_{ik}\delta_{jl} + \delta_{il}\delta_{jk}\right). \tag{23}$$

Here, $\delta$ is the Kronecker delta. $E$ is the modulus of elasticity and $\nu$ is the Poisson's ratio.

Again, assuming Perzyna's overstressed theory [16], we obtain $\dot{\varepsilon}_{ij}^{vp}$ in Equation (21) as:

$$\dot{\varepsilon}_{ij}^{vp} = \langle\Phi(F)\rangle\frac{\partial f_p}{\partial \sigma'_{ij}}; \ \langle\Phi(F)\rangle = \begin{cases}\Phi(F) & :F > 0 \\ 0 & :F \leq 0\end{cases}; F = \frac{f_l - f_r}{f_r}, \tag{24}$$

where $\Phi$ is the rate sensitivity function; $\langle \ \rangle$ is the Macaulay's bracket and $F$ is the overstress function. Additionally, $f_p$, $f_l$ and $f_r$ are the potential surface, the loading surface and the reference surface, which we discuss in the next section. It is worth noting that if $f_l < f_r$, the geomaterials behave elastically, while $f_l > f_r$, similar material will experience the viscoplastic strain. Additionally, we present the derivation of $\Phi$ in Appendix A. Moreover, we also discuss details derivation of $\dot{\varepsilon}_{ij}^{vp}$ for the finite element implementation in Appendix B.

3.2.2. Bounding Surfaces of the EVP Model

Constitutive models that adopt the classical plasticity theory, such as the MCC model, generally consider a single yield surface (SYF). The limitations of the SYF can be summarized as follows [15,18]:

(i) The SYF separates the elastic domain from the plastic state and forms an elastic state boundary within the yield surface. From the comparison of the experimental data and the model predictions, it is evident that the predicted elastic domain is larger than the observed one.

(ii) For the SYF models, the observed transition from the elastic state to the plastic state is in contrast to the experimentally observed gradual changes in the stiffness.

(iii) The SYF provides limited scopes to exemplify the plastic modulus in the loading direction.

(iv) The SYF model is usually incapable of capturing the proportional loadings.

During the last couple of decades, limitations of a single-surface model have opened up a more comprehensive research area. There are several methods to overcome these shortcomings. However, the two most popular theories are the multi-surface plasticity and the bounding surface model (see also Hashiguchi [18]; Yu [15]).

In this paper, for any loading history, we consider three bounding surfaces in the EVP model formulation (see also Equations (25) to (27) and Figure 2). Each surface has two ellipses: ellipse 1 (see also Kaliakin and Dafalias [26]) and ellipse 2 (see also Kutter and Sathialingam [27]). In the composite surface, two ellipses of each surface meet at common tangents and allow control of the shape of surfaces. In Figure 3, we illustrate surfaces where $\psi$ is the slope of the surface at any point on the potential surface. Additionally, as the deviatoric stress $(q)$ decreases with increases in the mean effective pressure $(p)$, the magnitude of $\psi$ is negative (see also Figure 3). To predict the overconsolidation effect of clay, the surface shape in the '*wet side*' is higher than the '*dry side*'. In this paper, assuming the ellipse shape parameter $(R)$ is equal to two, we obtain the extended MCC model equivalent ellipse shape (see also Islam et al. [14]).

$$\text{Potential surface, } f_p = \begin{cases} p_p^2 - \frac{2}{R}p_{cp}p_p - \frac{R-2}{R}p_{cp}^2 + (R-1)^2\left(\frac{q_p}{M}\right)^2 & :\text{Ellipse 1} \\ p_p^2 - \frac{2}{R}p_{cp}p_p + \left(\frac{q_p}{M}\right)^2 & :\text{Ellipse 2}\end{cases} \tag{25}$$

$$\text{Reference surface, } f_r = \begin{cases} p_r^2 - \frac{2}{R}p_{cr}p_r - \frac{R-2}{R}p_{cr}^2 + (R-1)^2\left(\frac{q_r}{M}\right)^2 & :\text{Ellipse 1} \\ p_r^2 - \frac{2}{R}p_{cr}p_r + \left(\frac{q_r}{M}\right)^2 & :\text{Ellipse 2}\end{cases} \tag{26}$$

$$\text{Loading surface, } f_l = \begin{cases} p_l^2 - \frac{2}{R}p_{cl}p_l - \frac{R-2}{R}p_{cl}^2 + (R-1)^2\left(\frac{q_l}{M}\right)^2 & :\text{Ellipse 1} \\ p_l^2 - \frac{2}{R}p_{cl}p_l + \left(\frac{q_l}{M}\right)^2 & :\text{Ellipse 2}\end{cases} \tag{27}$$

In Equations (25) to (27), the suffixes $p$, $r$ and $l$ represent the potential surface, the reference surface and the loading surface, respectively. $p_{cp}$, $p_{cr}$ and $p_{cl}$ are the intersection of the corresponding surface with the positive mean pressure axis (see also Figure 2). Additionally, $p = \frac{\acute{\sigma}_{kk}}{3}$ and $q = \left[\frac{3}{2}(\acute{\sigma}_d)_{ij}(\acute{\sigma}_d)_{ij}\right]^{\frac{1}{2}}$ are the mean effective normal stress and the deviatoric stress, respectively (see also Schofield and Wroth [21]). It is worth mentioning that we considered $p^t = \frac{\sigma_{kk}}{3}$ as the total mean stress. Also, $M$ represents the slope of the critical state line and is defined as (see also Prashant and Penumadu [28]):

$$M = \frac{6\sin\phi\sqrt{b^2-b+1}}{3+(2b-1)\sin\phi}, \tag{28}$$

where $\phi$ is the maximum internal friction angle for any specific stress path. $b$ ($0 \leq b \leq 1$) represents $b$-value (see also Islam and Gnanendran [23]; Islam et al. [14]). $b = 1$ and $b = 0$ designate the triaxial compression and the triaxial extension test, respectively. By changing the $b$-value, we obtain any stress path in the present model formulation. In Equations (25) to (27), we revise the expression of $M$ to obtain the realistic surface in the π-plane. Additionally, in Figure 4, we present a comparison of the MCC surface and proposed modification in the surfaces (see also Figure 2) with the true triaxial test results on Kaolin clay (see also Prashant and Penumadu [28]). Moreover, Lade [29] presented the relation between $b$-value and Lode angle as follows (see also Figure 5):

$$\cos(3\theta) = \frac{1}{2}\frac{(2b^3 - 3b^2 - 3b + 2)}{(b^2 - b + 1)^{\frac{3}{2}}}. \tag{29}$$

In Equation (29), $\theta$ and $b$ represent, the Lode angle and the $b$-value $\left(= \frac{\sigma_2-\sigma_3}{\sigma_1-\sigma_3} = \frac{\acute{\sigma}_2-\acute{\sigma}_3}{\acute{\sigma}_1-\acute{\sigma}_3}\right)$ respectively, while $\sigma_1$, $\sigma_2$ and $\sigma_3$ are the major principal stress, the intermediate principal stress and the minor principal stress, respectively, while $\acute{\sigma}_i$ represents their corresponding effective stress (see also Prashant and Penumadu [28]).

3.2.3. Image Parameters of the EVP Model

Using the triaxial test, we find the reference state for any clay and at any time. Additionally, the current loading stress state $(p_l, q_l)$ is associated with its image stress on the reference surface $(p_r, q_r)$ and the potential surface $(p_p, q_p)$ through the '*radial mapping rule*' (see also Hashiguchi [18]). Moreover, similar to the Modified Cam Clay (MCC) model, we also assume that the projection center is in the origin of the stress space (see also Figures 2 and 3). In this regard, Kutter and Sathialingam [27] reported that separation of the projection center and the stress state origin results in an elastic nucleus. Also, such a nucleus requires additional model parameters that split up the elastic domain from the inelastic domain (see also, Hashiguchi [18], Chapter 7). Therefore, ignoring the elastic nucleus, we obtain the image stress of the loading surface on the reference surface (see also Hashiguchi [18]) as follows:

$$\left.\begin{array}{rcl}\sigma_r &=& \beta_1\sigma_l\\ p_r &=& \beta_1 p_l\\ q_r &=& \beta_1 q_l\end{array}\right\}. \tag{30}$$

Now substituting Equation (30), into Equation (26) for Ellipse 1 ($\eta < M$) (see also Figure 3):

$$\beta_1 = \frac{\frac{2p_l p_{cr}}{R} + \frac{2p_l p_{cr}}{R}\sqrt{1+\left[1+\frac{(R-1)^2}{p_l^2}\left(\frac{q_l}{M}\right)^2\right](R-2)R}}{2\left[p_l^2+(R-1)^2\left(\frac{q_l}{M}\right)^2\right]}. \tag{31}$$

Similarly, for $\eta > M$ and Ellipse 2, we find:

$$\beta_1 = \frac{2p_{cr}}{R\left[\left(p_l+\frac{1}{p_l}\left(\frac{q_l}{M}\right)^2\right)\right]}. \tag{32}$$

Again, for the potential surface, we also find:

$$\beta_2 = \frac{\frac{2p_l p_{cp}}{R} + \frac{2p_l p_{cp}}{R}\sqrt{1+\left[1+\frac{(R-1)^2}{p_l^2}\left(\frac{q_l}{M}\right)^2\right](R-2)R}}{2\left[p_l^2+(R-1)^2\left(\frac{q_l}{M}\right)^2\right]}. \tag{33}$$

$$\beta_2 = \frac{2p_{cp}}{R\left[\left(p_l+\frac{1}{p_l}\left(\frac{q_l}{M}\right)^2\right)\right]}. \tag{34}$$

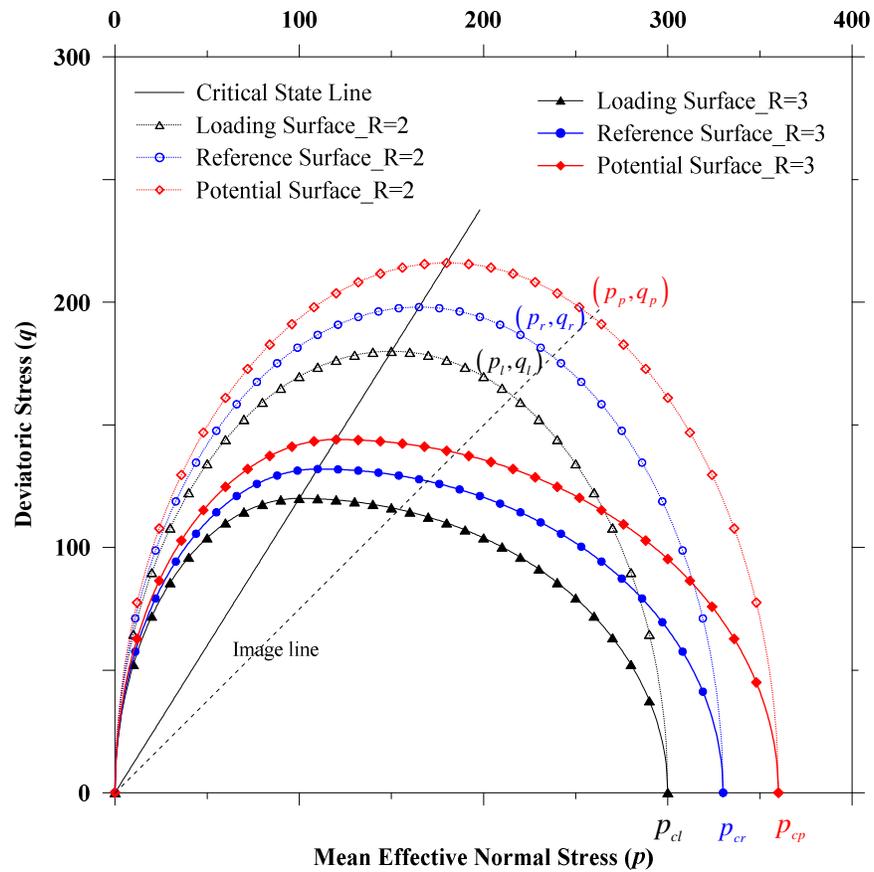

**Figure 2.** In the *p*-*q* plane, schematic representation of the potential surface, the reference surface and the loading surface and the surface shape parameter effect.

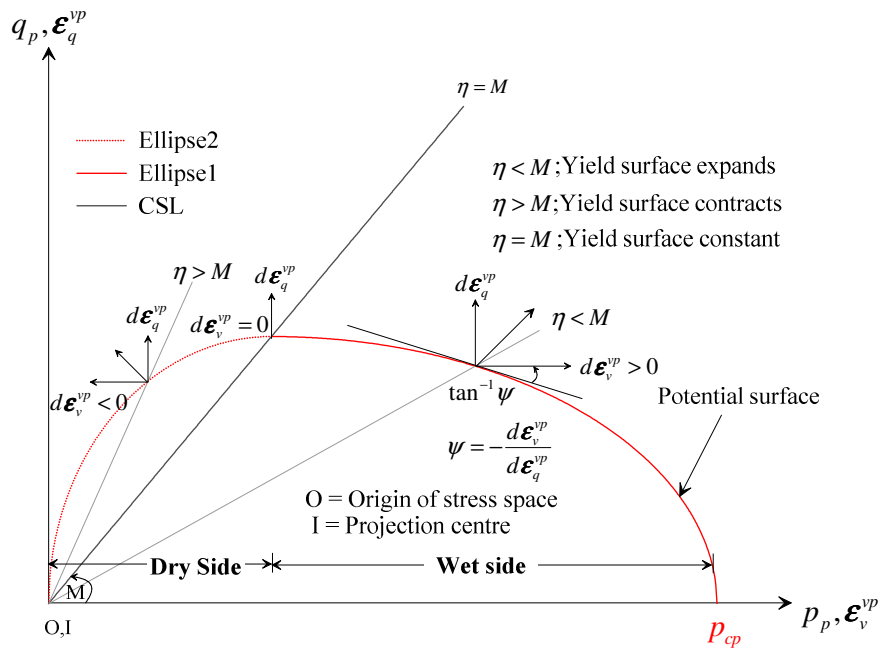

**Figure 3.** Meridional section of the potential surface with ellipse 1 and ellipse 2.

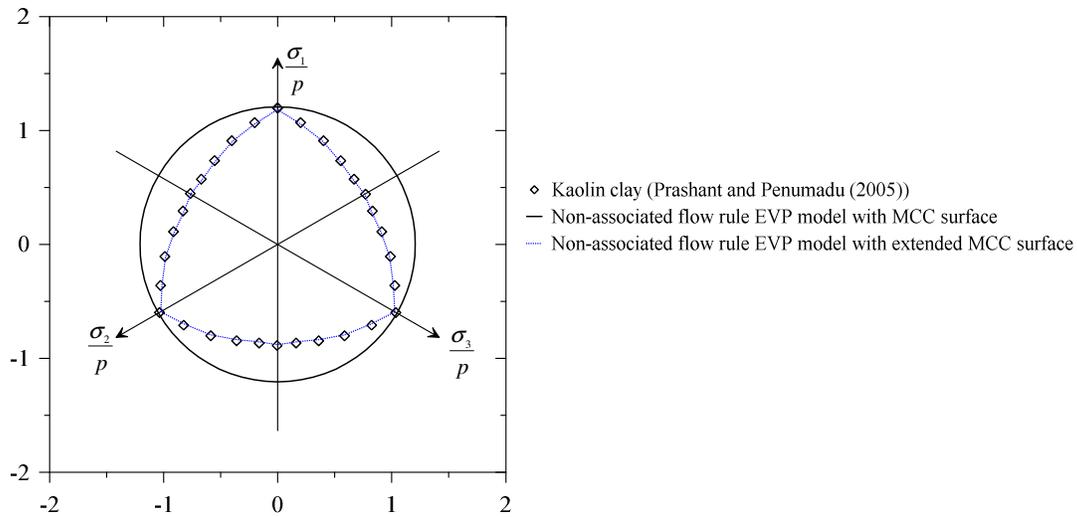

**Figure 4.** In the π-plane, comparison of the true triaxial test (TTT) results with the original Modified Cam Clay (MCC) surface and the extended MCC surface (for TTT test results see also Prashant and Penumadu [28]).

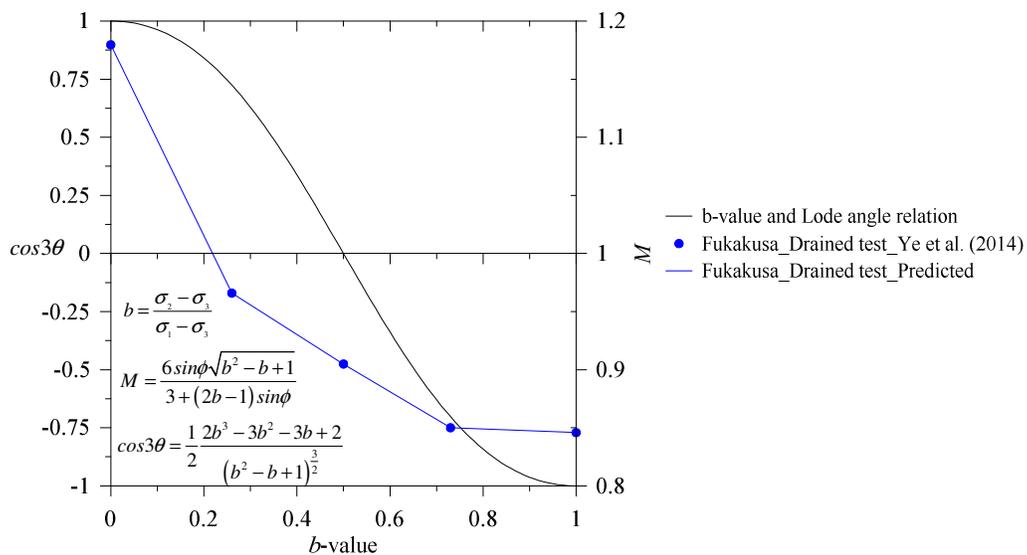

**Figure 5.** Relations of *b*-value-Lode Angle and *b*-value-*M* (see also Lade [29]; Islam and Gnanendran [23]; Ye et al. [30]).

It is noteworthy that in Equations (31) to (34) for the Modified Cam Clay (MCC) equivalent surface (R = 2, see also Figure 2), we obtain $\beta_1 = \frac{p_{cr}}{p_{cl}}$ and $\beta_2 = \frac{p_{cp}}{p_{cl}}$ (see also Islam et al. [14]), where, $p_{cl} = p_l + \frac{1}{p_l}\left(\frac{q_l}{M}\right)^2$. In Appendix A, we discuss the derivation of $p_{cr}$ and $p_{cp}$ (see also Figure 2).

### 3.3. Finite Element Implementation

#### 3.3.1. Couple Finite Element Formulation

We assume that a soil mass occupies in a domain $\Omega$ and its surface is $\Gamma$, which is subdivided into a subdomain and segment of the surface during the discretization of space, as $\partial\Omega$ and $\partial\Gamma$, respectively. In the following sequences, to obtain a coupled solution for the governing equations (see Equations (9) to (11), (20)), we use the weak form solution and the Galerkin weighted residual method (see also Zienkiewicz et al. [31]). We use the effective stress relation (see Equation (12)) and the Darcy's law (see Equation (16)) to obtain the coupled hydro-mechanical relationship for the elasto-viscoplastic model. In our model formulation, we have a total of 16 equations. One equation for the equilibrium relation (see Equation (9)). Three equations for the mass balance relations (see Equations (10) and (11)). Moreover, we present six equations for the stress-strain relationship (see Equations (22) and (24)) and six equations for the strain displacement relation (see Equation (19)). Additionally, we have 16 unknowns (six for the effective stress, six for the strain, three for the displacement and one for the liquid pressure). Thereby, our coupled solutions represent a "well-posed" problem definition (see also Bear and Bachmat [24])

Additionally, the element matrices for two phases porous media are given by (see also Zienkiewicz et al. [31]):

$$\begin{bmatrix} K_E & I_E \\ [I_E]^T & 0 \end{bmatrix} \begin{bmatrix} \dot{u}^n \\ \dot{p}_f^n \end{bmatrix} + \begin{bmatrix} 0 & 0 \\ 0 & H_E \end{bmatrix} \begin{bmatrix} u^n \\ p_f^n \end{bmatrix} = \begin{bmatrix} F_E \\ Q_E \end{bmatrix}. \tag{35}$$

We also find the global matrix by a sum over the number of elements in the element matrix. Also, in a simplified form, we re-write Equation (35) as (see also Owen and Hinton [32]):

$$C\dot{X} + KX = F \tag{36}$$

where $C = \begin{bmatrix} K_E & I_E \\ [I_E]^T & 0 \end{bmatrix}$, $K = \begin{bmatrix} 0 & 0 \\ 0 & H_E \end{bmatrix}$, $F = \begin{bmatrix} F_E \\ Q_E \end{bmatrix}$, $\dot{X} = \begin{bmatrix} \dot{u}^n \\ \dot{p}_f^n \end{bmatrix}$ and $X = \begin{bmatrix} u^n \\ p_f^n \end{bmatrix}$,

$$K_E = \int_\Omega [B_u]^T D B_u \, d\Omega, \tag{37}$$

$$I_E = \int_\Omega [B_u]^T m N_p \, d\Omega, \tag{38}$$

$$[I_E]^T = \int_\Omega [N_p]^T m^T B_u \, d\Omega, \tag{39}$$

$$H_E = -\int_\Omega [B_p]^T \frac{K}{\gamma_w} B_p \, d\Omega, \tag{40}$$

$$F_E = \int_\Omega [N_u]^T b \, d\Omega + \oint_\Gamma [N_u]^T \dot{T} \, d\Gamma + \int_\Omega [B_u]^T D \dot{\varepsilon}^{vp} \, d\Omega, \tag{41}$$

$$Q_E = -\int_\Omega [B_p]^T \frac{K}{\gamma_w} b_f \, d\Omega - \oint_\Gamma [N_p]^T w \, d\Gamma, \tag{42}$$

$$D = [(D^e)^{-1} + C]^{-1}, \tag{43}$$

$$C = \theta_d \Delta t_n H_n, \tag{44}$$

$$H_n = \left(\frac{\partial \dot{\varepsilon}^{vp}}{\partial \sigma'}\right)_n, \tag{45}$$

$$m = [1,1,1,0,0,0]^T, \tag{46}$$

$$N_u = \begin{bmatrix} N_{u1} & 0 & 0 & \ldots & N_{un} & 0 & 0 \\ 0 & N_{u1} & 0 & \ldots & 0 & N_{un} & 0 \\ 0 & 0 & N_{u1} & \ldots & 0 & 0 & N_{un} \end{bmatrix}. \tag{47}$$

$$N_p = [N_{p1} \quad \ldots \quad N_{pn}] \tag{48}$$

In the above equations, $K_E$ is the tangential stiffness matrix. $I_E$ is the coupling matrix. $F_E$ is the load vector. $H_E$ is the flux matrix and $Q_E$ is the fluid conduction matrix. $T$ and $T$ represent transpose and the traction force, respectively. $D^e$ is the elastic constitutive matrix related to $S$ (see also Equations (22) and (23)). $B_u = \mathcal{L} N_u$ is the strain-displacement matrix, where $\mathcal{L}$ is the tangential operator. $N_u$ and $N_p$ are the displacement shape function and the pore pressure shape function, respectively. $t_n$ represents time. $\theta_d$ demonstrates the integration parameter (see also Segerlind [33]). $m$ is a mapping vector. $H_n$ is the gradient matrix (see also Owen and Hinton [32]).

It is worth mentioning that for the isoparametric element, the shape function and the interpolation function are identical (see Zienkiewicz et al. [31]; Potts and Zdravkovic [34]). This simplification allows flexibility to consider any arbitrary shape of elements. Additionally, the shape functions are defined in terms of the local coordinates $(\xi, \eta, \zeta)$ and the strain interpolation matrix requires global derivatives, with respect to the global coordinates $(x, y, z)$. To map both coordinates, the chain rule can be applied as follows (see also Zienkiewicz et al. [31]):

$$\begin{bmatrix} \frac{\partial \bar{N}_i}{\partial \xi} \\ \frac{\partial \bar{N}_i}{\partial \eta} \\ \frac{\partial \bar{N}_i}{\partial \zeta} \end{bmatrix} = \begin{bmatrix} \frac{\partial x}{\partial \xi} & \frac{\partial y}{\partial \xi} & \frac{\partial z}{\partial \xi} \\ \frac{\partial x}{\partial \eta} & \frac{\partial y}{\partial \eta} & \frac{\partial z}{\partial \eta} \\ \frac{\partial x}{\partial \zeta} & \frac{\partial y}{\partial \zeta} & \frac{\partial z}{\partial \zeta} \end{bmatrix} \begin{bmatrix} \frac{\partial \bar{N}_i}{\partial x} \\ \frac{\partial \bar{N}_i}{\partial y} \\ \frac{\partial \bar{N}_i}{\partial z} \end{bmatrix} = [J] \begin{bmatrix} \frac{\partial \bar{N}_i}{\partial x} \\ \frac{\partial \bar{N}_i}{\partial y} \\ \frac{\partial \bar{N}_i}{\partial z} \end{bmatrix}. \qquad (49)$$

In Equation (49), $[J]$ is the Jacobian matrix. $\bar{N}_i$ is the shape function for nodal values, where $i$ represents nodal points of elements. In the literature, there are two types of assumptions to account for the shape function for the displacement and the pore pressure. In the first case, it is assumed that both shape functions are identical, while in the second case, different shape functions are adopted for variables groups. For example, the displacement and the pore pressure variables vary linearly in the linear triangular element (e.g., three nodes) and the bilinear rectangular element (e.g., four nodes) (Zienkiewicz et al. [31]; Potts and Zdravkovic [34]). However, in a six nodes triangular element and an eight nodes rectangular element, the displacement and the pore pressure field change quadratically. Additionally, when the displacement varies quadratically, the effective stress change linearly (see also Zienkiewicz et al. [31]). Thereby, there is a variation between the pore pressure and effective stress. To achieve the same order of variation in the primary variables, for the eight nodes rectangular element, the degrees of freedom (DOF) for the pore pressure can be obtained at four corners of the rectangular element. In contrast, for the triangular element, a similar DOF for the pore pressure is calculated only from the apex of the triangle (see Potts and Zdravkovic [34]). Hence, the shape function for the displacement and the pore pressure can be separated. Moreover, to account for the large deformation for the homogeneous porous media (e.g., triaxial creep or relaxation test) and heterogeneous porous media (e.g., embankment with multiple layers), special attention is essential to consider the element type (see also Zienkiewicz et al. [31]). For example, to model the consolidation behavior of porous media comprising sand deposit overlying clay deposit, the sand is modeled assuming no pore pressure DOF at the nodes, which behaves like drained media or non-consolidating elements. In contrast, clay deposit is modeled as a consolidating element considering fluid pressure degrees of freedom at the nodes (see Potts and Zdravkovic [34]). Additionally, Zienkiewicz et al. [31] demonstrated that when porous media is approaching the undrained limit state, to satisfy the Babuska-Brezzi convergence condition (see Zienkiewicz et al. [31]; Potts and Zdravkovic [34]), the shape function for the nodal displacement and the pore pressure need to be separated. In such a case, the choice of element types is limited, and details can be found in Zienkiewicz et al. [31]. In any other state than the undrained limit state, element type selection is extensive in the finite element simulation practice.

In the *Results and Discussion* Section, we discuss both the drained and the undrained triaxial tests considering the short term loading and the long-term loading. Additionally, in this paper, for validation of the triaxial test, we use the first-order three nodes triangular element. Moreover, for the long-term prediction (e.g., embankment performance estimation), we consider the second-order six nodes triangular element (see also Zienkiewicz et al. [31]).

3.3.2. Time Integration

For finite element modeling (FEM) of time-dependent porous media (e.g., clay), there are several approaches to discretize the time domain (see also Segerlind [33])). However, for the FEM solutions, the $\theta$-method is the simplest method (see Potts and Zdravkovic [34]) and in this paper we also use this method. The value of $\theta$ ranged in between 0 to 1 while $\theta = 0, 0.5, 1$ represent the fully explicit time integration (also known as the forward difference method), the implicit trapezoidal time integration (also known as the Crank-Nikolson rule) and the fully implicit time integration (also known as the backward difference method), respectively (see also Zienkiewicz et al. [31]; Owen and Hinton [32]; Segerlind [33]; Potts and Zdravkovic [34]).

Again, nodal values of the interpolation polynomial $(N_i)$ for variables (e.g., displacement, pore pressure) at nodal points varies with respect to individual element's nodal coordinates. For example, for the linear triangular element $N_i$ has three nodal points. Segerlind [33] presented, Equation (36) for the interpolation polynomial of the nodal values as follows:

$$\boldsymbol{C}\dot{\boldsymbol{N}}_i + \boldsymbol{K}\boldsymbol{N}_i = \boldsymbol{F}. \tag{50}$$

From Figure 6, for a given function $N_i$ with time interval $\Delta t = t_{n+1} - t_n$, using the mean value theorem, we obtain:

$$\frac{dN_i}{dt}(Y) = \frac{N_i(t_{n+1}) - N_i(t_n)}{\Delta t}, \tag{51}$$

where $Y$ represents time (see also Figure 6 and Segerlind [33])

Additionally, from Figure 6 and Equation (51), the value of $N_i(Y)$ can be obtained as:

$$N_i(Y) = N_i(t_n) + (Y - t_n)\frac{N_i(t_{n+1}) - N_i(t_n)}{\Delta t}. \tag{52}$$

Assuming, the integration parameter $\theta = \frac{(Y - t_n)}{\Delta t}, \theta \in 0,1$ Equation (52) becomes:

$$N_i(Y) = (1 - \theta)N_i(t_n) + \theta N_i(t_{n+1}). \tag{53}$$

Again, at $t = Y$ (see Figure 6), $\boldsymbol{F}$ in Equation (50) is given by (see Segerlind [33]):

$$\boldsymbol{F} = (1 - \theta)\boldsymbol{F}(t_n) + \theta \boldsymbol{F}(t_{n+1}). \tag{54}$$

Substituting Equations (51) to (54), into Equation (50), we obtain:

$$[\boldsymbol{C} + \theta \Delta t \boldsymbol{K}]N_i(t_{n+1}) = [\boldsymbol{C} - (1 - \theta)\Delta t \boldsymbol{K}]N_i(t_n) + \Delta t[(1 - \theta)\boldsymbol{F}(t_n) + \theta \boldsymbol{F}(t_{n+1})]. \tag{55}$$

By changing $N_i$ with corresponding primary variables, $\boldsymbol{X} = \begin{bmatrix} \boldsymbol{u}^n \\ \boldsymbol{p}_f^n \end{bmatrix}$ (see Equation (36)), we obtain a coupled two-phase solution.

For $\theta = 0, 0.5$ and 1, we also find a simplified form of Equation (55) (see also Segerlind [33]).

Additionally, to demonstrate the effect of $\Delta t$ and $\theta$ in solutions, for simplicity, we assume an ordinary differential equation as follows:

$$\begin{aligned} \acute{y} &= -y, \\ y(0) &= 1, \\ 0 &\leq t \leq T, \\ \Delta t &= \frac{T}{N}, \end{aligned} \tag{56}$$

where $T$ and $N$ are the total time and time step incremental number, respectively. The exact solution of Equation (56) is given by:

$$y(t) = e^{-t}. \tag{57}$$

Moreover, we also find the solution of Equation (57) assuming the backward Euler or implicit Euler and the forward Euler or the explicit Euler as follows, respectively:

$$y_n = \frac{1}{(1 + \Delta t)^n}, \tag{58}$$

$$y_n = (1 - \Delta t)^n. \tag{59}$$

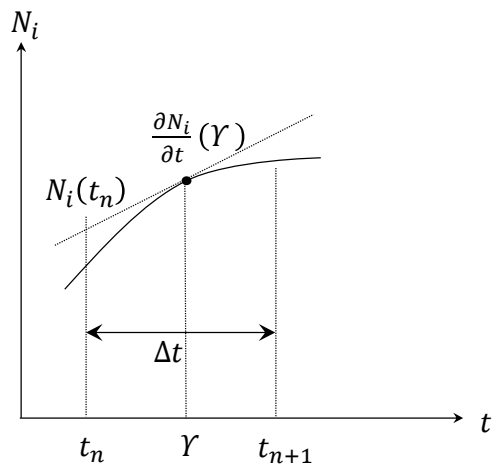

**Figure 6.** An approximation of $N_i$ for given time increment $\Delta t = t_{(n+1)} - t_n$.

From Equations (58) and (59), we find that for any value $\Delta t > 1$, there will be notable oscillation in the forward Euler or the explicit Euler solution. In Figure 7, we present the effect of the selection of $\Delta t$ and $\theta$ in solutions.

Again, the solution of the coupled Equation (55) for the non-associated flow rule-based EVP model herein is more complicated compared to Equation (57). However, the solution scheme of Equation (55) holds a similar degree of convergence challenge to select $\Delta t$ and $\theta$ as that evident in Figure 7. Thereby, for the solution of the hydro-mechanical coupled equation, careful consideration is essential to choose $\Delta t$ for the corresponding $\theta$-method (see also Owen and Hinton [32]). For our solution, we assume $\theta = 0.5$ and we obtain critical values of $\Delta t$ following Potts and Zdravkovic [34].

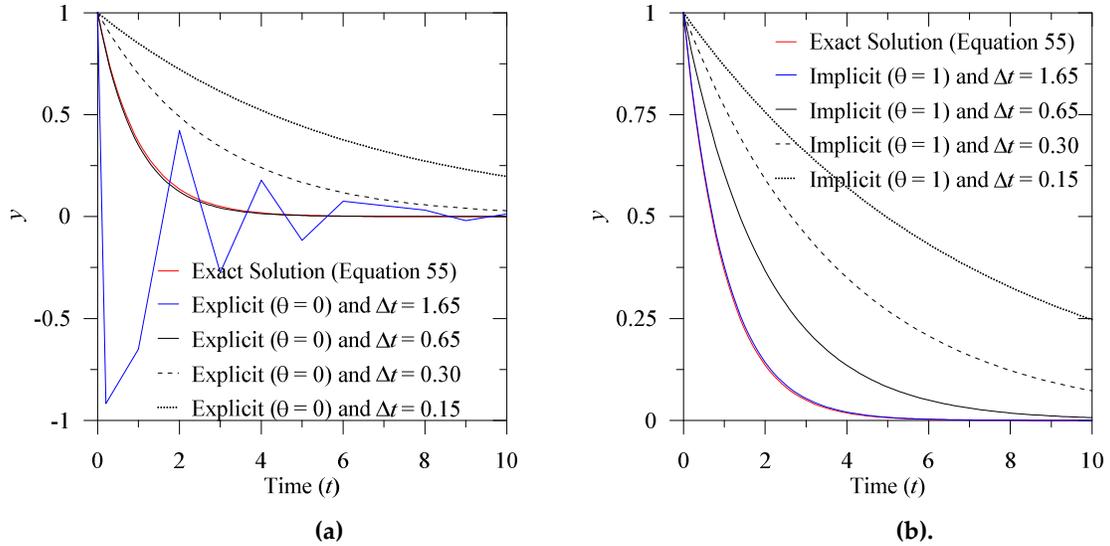

**Figure 7.** Effect of $\Delta t$ for the solution using $\theta$-method (**a**) Explicit and (**b**) Implicit.

3.3.3. Incremental Stress and Strain

In Equation (55), replacing $N_i$ with the viscoplastic strain rate $(\dot{\varepsilon}_{ij}^{vp})$, we obtain:

$$\Delta \boldsymbol{\varepsilon}_n^{vp} = \Delta t\big[(1-\theta)\dot{\boldsymbol{\varepsilon}}_n^{vp} + \theta \dot{\boldsymbol{\varepsilon}}_{n+1}^{vp}\big]. \tag{60}$$

Again, using the Taylor series expansion and ignoring the higher-order of $\dot{\boldsymbol{\varepsilon}}_{n+1}^{vp}$, Equation (60) is given by (see also Owen and Hinton [32]):

$$\dot{\boldsymbol{\varepsilon}}_{n+1}^{vp} = \dot{\boldsymbol{\varepsilon}}_n^{vp} + \boldsymbol{H}_n \Delta \acute{\boldsymbol{\sigma}}_n. \tag{61}$$

We presented $\boldsymbol{H}_n$ in Equation (45). Now, substituting Equation (61) into Equation (60), we obtain:

$$\Delta \boldsymbol{\varepsilon}_n^{vp} = \Delta t \dot{\boldsymbol{\varepsilon}}_n^{vp} + \Delta t \Delta \theta_d \boldsymbol{H}_n \acute{\boldsymbol{\sigma}}_n. \tag{62}$$

Combining Equations (22) and (24), the incremental stress can be rewritten as (see also Owen and Hinton [32], p. 272):

$$\Delta \acute{\boldsymbol{\sigma}}_n = \boldsymbol{D}\big(\Delta \boldsymbol{\varepsilon}_n - \Delta t_n \dot{\boldsymbol{\varepsilon}}_n^{vp}\big), \tag{63}$$

where, the total incremental strain, $\Delta \boldsymbol{\varepsilon}_n = \boldsymbol{B}^n \Delta \boldsymbol{u}^n$ and $\Delta \boldsymbol{u}^n$ is the incremental displacement, while $\boldsymbol{B}$ is the strain-displacement matrix. Additionally, from Equations (61) and (62), we find $\dot{\boldsymbol{\varepsilon}}_n^{vp}$ in Equation (63). Moreover, we obtain $\boldsymbol{D}$ from Equation (43) (see also Owen and Hinton [32], p.274).

*3.4. Initial and Boundary Conditions*

For two-phase coupled porous mediums, we find initial conditions of primary variables as follows:

$$\boldsymbol{u}(t=0) = \boldsymbol{u}_0 \qquad \text{in } \Omega, \tag{64}$$

$$p_l(t=0) = p_{l_0} \quad \text{in } \Omega. \tag{65}$$

Additionally, the boundary conditions are given by:

$$\boldsymbol{\sigma} \cdot \boldsymbol{n} = \boldsymbol{t} \quad \text{on } \Gamma_t, \tag{66}$$

$$\boldsymbol{u} = \boldsymbol{u}_0 \quad \text{on } \Gamma_u, \tag{67}$$

$$p_l = p_{l_0} \quad \text{on } \Gamma_{p_{l'}} \tag{68}$$

$$\boldsymbol{v} \cdot \boldsymbol{n} = 0 \quad \text{on } \Gamma_{v^l}, \tag{69}$$

where, $\boldsymbol{u}_0$ and $p_{l_0}$ are the initial displacement and the initial porewater pressure. $\boldsymbol{n}$ represents the unit normal vector. $\boldsymbol{t}$ denotes the traction force. We have two types of boundary conditions. They are the Neumann boundary conditions (e.g., $\Gamma_t$) and the Dirichlet boundary condition (e.g., $\Gamma_u$).

In this paper, for the validation of the non-associated flow rule-based elasto-viscoplastic model, we use the conventional triaxial test (e.g., cylindrical specimen). We discuss the details of the validation in the *Results and Discussion* Section. For validation, we consider different clay samples. Thereby, the initial conditions of those triaxial samples are also different, which we obtain from the published literature. Depending on the clay sample, the initial stress state (e.g., $\sigma_i$), the initial pore pressure $(p_{l_0})$, the loading rate (e.g., stress-controlled or strain-controlled test) and the confining pressure are different. We assume that in the initial state, all clay samples are fully saturated. Additionally, we consider an axisymmetric section of the triaxial sample for the EVP model validation. In Figure 8, we present a representative illustration of the initial and the boundary condition of the conventional triaxial sample.

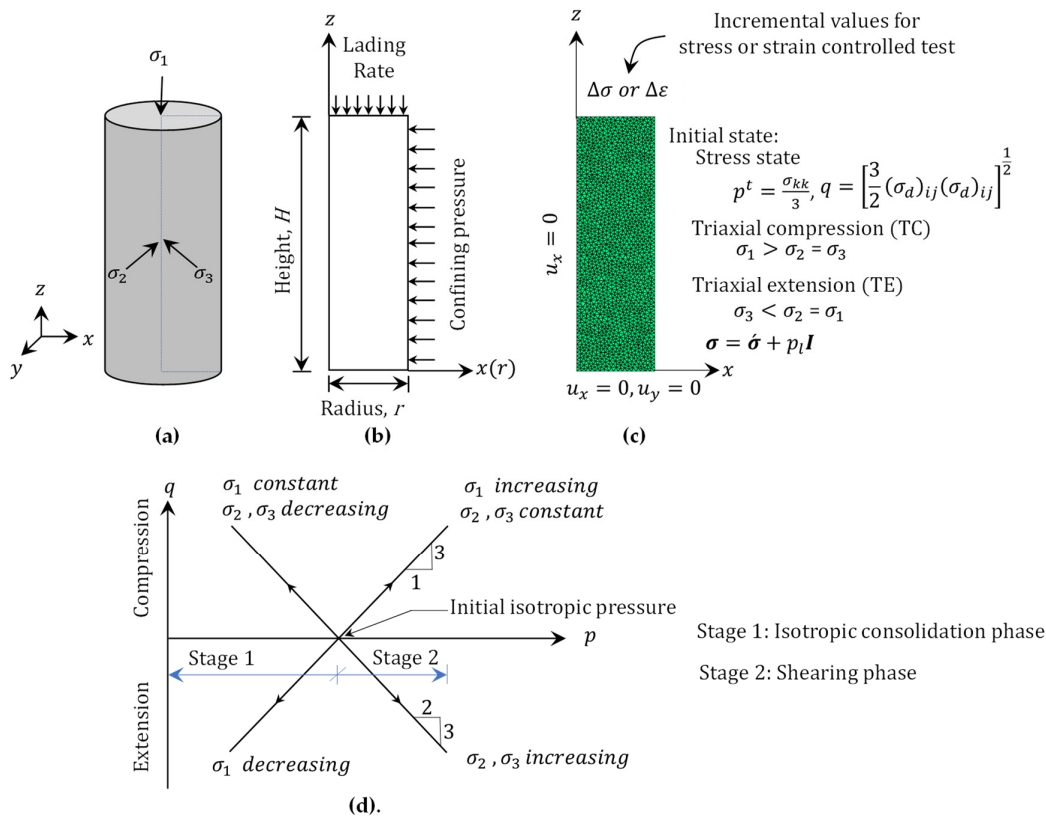

**Figure 8.** The Cambridge stress state in a conventional triaxial test (**a**) a cylindrical sample, (**b**) an axisymmetric section, (**c**) a finite element mesh and (**d**) p-q diagram for loading state (see also Lade [29]), p. 62).

Again, both the drained and the undrained tests consist of two stages. They are the isotropic consolidation stage and the shearing stage (see also Figure 8d). Additionally, for both tests, the isotropic consolidation procedure is identical. During the first stage of the triaxial test, on the bottom, we assume that the horizontal and the vertical displacement components are fixed. Additionally, due to the symmetry of the axis, we also restrict the horizontal displacement along the radial or *x*-axis in the left side boundary. We also apply the confining pressure along the right boundary. In addition, depending on the stress-controlled or the strain-controlled test, we also introduce the corresponding value at the top. Moreover, in the initial state, we allow the drainage at the top boundary. Also, we invoke the "geostatic" stress condition considering the body force of the sample. The "geostatic" condition ensures the equilibrium state of the sample. Furthermore, we consider the stress ratio along the horizontal direction to the vertical direction is 1. Then, we run the axisymmetric triaxial sample for one incremental step, which we assume as the initial state for the shearing stage. The initial state makes sure that the triaxial sample satisfies the initial yield surface of the elasto-viscoplastic model.

Additionally, in the shearing stage, there are two essential criteria for the drained and the undrained tests. The first one is the boundary condition, and the latter one is the loading rate. For the drained test, the top and the bottom boundaries are permeable, while these boundaries are impermeable for the undrained test. Additionally, for the drained test, the loading rate is slow to avoid the development of the excess pore water pressure, while in the undrained test, the loading rate is fast. It is noteworthy that for the drained condition, selection of the time increment is crucial, and it needs to be small (see also Figure 7). After the selection of the optimal time step for the drained test, we plot the excess pore water (EPW) pressure with respect to time, confirming that the EPW is approximately zero. Moreover, in the undrained test, immediately after the incremental load application, the excess pore water pressure builds up. Therefore, initially, we consider a small-time step for the undrained test, and then we increase the time step incrementally.

In Figure 8d, we discuss two stress paths for the cylindrical sample. They are the triaxial compression (TC) and the triaxial extension (TE). After the isotropic consolidation state, we apply $\sigma_1$, $\sigma_2$ and $\sigma_3$ for corresponding TC ad TE tests. It is worth mentioning that the EVP model presented in this paper is not limited to any specific geometry shape or the specific stress paths, which are in most cases, the limitations of many elasto-viscoplastic models in the literature. In the present paper, by changing the *b*-value, we will able to obtain any stress path in the stress space. Additionally, we will able to introduce a specialized stress path too. For example, the constant mean pressure test or the constant $\sigma_3$ test (see also Figures 4, 5 and Lade [29]). For such a special loading condition, we need to revise the corresponding $\sigma_1$, $\sigma_2$ and $\sigma_3$ with respect to the mean pressure, the deviatoric pressure and the *b*-value. From three equations, we obtain three unknowns, $\sigma_1$, $\sigma_2$ and $\sigma_3$.

*3.5. Model Parameters*

In this paper, the non-associated flow rule-based elasto-viscoplastic model requires a total of seven parameters. They are divided into (i) consolidation parameters, (ii) strength parameter, (iii) elastic property, (iv) state parameter, (v) creep parameter and (vi) surface shape parameter. The consolidation properties are the normal consolidation line gradient $\left(\lambda = \frac{C_c}{2.3}\right)$ and the swelling line gradient $\left(\kappa = \frac{C_s}{2.3}\right)$, where $C_c$ and $C_s$ are the compression index and the swelling index, respectively (see also Figure A1 in Appendix A). The slope of the critical state line $(M)$ is considered as the strength parameter and related to the angle of internal friction at failure (see Figures 2, 5 and Equation (28)). We consider the Poisson's ratio $(\nu)$ as the elastic property. Additionally, at any reference time and unit mean pressure, the void ratio $(e_N)$ is the state parameter (see also Figure A1 in Appendix A). Moreover, to account for the time-dependent behavior of clay, we introduce the secondary compression index $(C_\alpha)$ as the creep parameter. Finally, we also introduce a surface shape parameter $(R)$ to control the shape of the bounding surface (see Figures 2 and 3 and Equations (25) to (27)). In Table 1, we present a summary of the non-associated flow rule-based elasto-viscoplastic (EVP) model parameters in this paper and their determination method.

**Table 1.** Model parameters in the elasto-viscoplastic model.

| Parameters | Meaning | Method of determination |
|---|---|---|
| $\lambda$ | Slope of NCL | Triaxial or oedometer test |
| $\kappa$ | Slope of SL | Triaxial or oedometer test |
| M | Slope of CSL | Triaxial test |
| $\nu$ | Poisson's ratio | Assumed |
| $e_N$ | Void ratio at $p = 1$ with NCL at $\bar{t}$ | Triaxial or oedometer test |
| $C_\alpha$ | Creep parameter | Triaxial or oedometer test |
| $R$ | Shape parameter | Undrained triaxial test |

Note: NCL = the normal consolidation line, SL = the swelling line, CSL = the critical state line, $\bar{t}$ is the reference time. Also, $E = \frac{3(1-2\nu)(1+e_0)p}{\kappa}$ (see also Islam et al. [14]), $G = \frac{E}{2(1+\nu)}$

It is noteworthy that among seven parameters, five parameters are identical to the Modified Cam Clay (MCC) model. The other two parameters are $C_\alpha$ and $R$. Additionally, the MCC model parameters determination processes are well documented in many soil mechanics textbooks (see also Schofield and Wroth [21]; Roscoe and Burland [10]). Moreover, there are three approaches available in the literature to obtain $C_\alpha$. The first one is the empirical relation with $C_c$ (see also Mesri and Castro [35]) which is assumed constant over time. The second one is a direct calculation from the oedometer test or the triaxial test. This approach is also divided into the void ratio-based method $\left(C_{\alpha e} = \frac{\Delta e}{\Delta \log t}\right)$ and the strain-based method $\left(C_{\alpha \varepsilon} = \frac{\Delta \varepsilon}{\Delta \log t} = \frac{C_{\alpha e}}{1+e_0}\right)$ (see also, Liingaard et al. [11]). Additionally, using the cone-penetration test field data, we also may obtain $C_\alpha$ (see also Tonni and Simonini [36]). Moreover, in the literature, there are two concepts regarding the secondary compression index: constant or linear function and non-linear function, while in both cases, strong opinions are available. In this paper, we use the void ratio-based non-linear $C_\alpha$ which is neither tied to any specific clay nor requires any fittings parameters (see also Islam and Gnanendran [23]; Islam et al. [14]) and reads:

$$\frac{C_{\alpha_i}}{C_{\alpha_{i-1}}} = \left(\frac{p_{cp_i}}{p_{cp_{i-1}}}\right)^{\frac{\lambda-\kappa}{\alpha_{i-1}}}, \tag{70}$$

where $i$ and $(i-1)$ represent the present step and the previous time step, respectively.

Additionally, to obtain the surface shape function $(R)$, there are several methods. For example, Dafalias and Herrmann [37] proposed a fitting method using the undrained triaxial stress path as follows:

$$\frac{|q_p|}{p_{cp}} = \frac{M}{R-1}\left[\frac{2}{R}\left(\frac{p_{cp}}{p_{cp}}\right)^{\frac{\lambda-2\kappa}{\lambda-\kappa}} + \left(1-\frac{2}{R}\right)\left(\frac{p_p}{p_{cp}}\right)^{\frac{-2\kappa}{\lambda-\kappa}} - \left(\frac{p_p}{p_{cp}}\right)^2\right]^{\frac{1}{2}}. \tag{71}$$

It is worth mentioning that to predict the pore pressure for the undrained test, we also need the permeability (see Equation (16)), which is the material parameter of clay and not relevant to the EVP model parameters. Additionally, in Table 2, we summarize EVP model parameters for different clays.

Table 2. Elasto-viscoplastic (EVP) model parameters for different clays.

| Clay | EVP Model Parameters in This Paper | | | | | | | |
|---|---|---|---|---|---|---|---|---|
| | $\lambda$ | $\kappa$ | $M_c$ | $M_e$ | $\nu$ | $e_N$ | $C_\alpha$ | $R$ |
| Shanghai clay [38] | 0.22 | 0.046 | 1.28 | --- | 0.30 | 2.23 | 0.016 | 2.00 |
| SFBM clay [39] | 0.37 | 0.054 | 1.40 | --- | G | 3.17 | 0.053 | 2.10 |
| Kaolin clay [40] | 0.15 | 0.018 | 1.25 | 0.95 | 0.30 | 1.51 | 0.014 | 2.50 |

Note: SFBM = San Francisco Bay Mud clay; G = 23540 kPa (see Kaliakin and Dafalias [26]); *Mc* and *Me* represent the triaxial compression and triaxial extension test, respectively.

## 4. Results and Discussion

*4.1. Shanghai Clay*

From Huang et al. [38], we obtain the Shanghai soft clay properties and model parameters. It is an undisturbed soft sensitive clay (sensitivity = 4.86). The water content, the liquid limit and the plastic limit of this clay sample are 51.80%, 44.17% and 22.40%, respectively. Additionally, the specific gravity of the Shanghai clay is 2.74. Moreover, the clay fraction ($< 5\ \mu m$), the silt fraction and the sand fraction of this clay are 26.60%, 63.40% and 10.00%, respectively. Furthermore, Huang et al. (2011) demonstrated that diameter and length of triaxial samples were 39.10 mm and 80.00 mm, respectively, which we use for the preparation of the numerical model geometry.

In Figures 9, we compare numerical results with laboratory-measured stress controlled, isotropic and consolidated undrained triaxial tests (see also Huang et al. [38]). Additionally, we also compare the EVP model predictions with the Kutter and Sathialingam [27] proposed EVP model. From the comparison, we find that in the small strain zone and after 14% axial strain, both the EVP models over predict the deviatoric stress. It is to note that in the EVP model formulation, we ignore the hysteretic response of the clay (see also Whittle and Kavvadas [41]), which results in the overprediction in the early stage. Also, in the present model formulation, we did not include the destructured behavior of clay (see also Liu and Carter [42]). Therefore, we observe a slight overprediction at the higher axial strain. Inevitably, in the present EVP model formulation, the incorporation of the hysteretic response and the destructured response require extra model parameters.

In Figure 9a, we compare the associated flow rule and the non-associated flow rule EVP models predicted stress-strain responses with the Shanghai clay observed results. We find that for 150 kPa mean pressure up to 1% axial strain; both models capture experimental responses well. Then, before the measured peak deviatoric stress, we also observe overprediction. Afterward, under-prediction in the associated flow rule-based EVP model continues. We also observe similar results for 200 kPa. Comparing both the flow rule EVP model and the Kutter and Sathialingam [27] model, we also find that the non-associated flow rule model well captures the experimental results.

Furthermore, we present the stress path responses in Figure 9b. We observe that after attaining the peak deviatoric stress, the EVP model prediction gradually follows a 'narrow region' (see also Islam et al. [14])) and such a phenomenon is common in the natural clay than the reconstituted clay (see also Islam and Gnanendran [23]). In general, the MCC model (Roscoe and Burland [10]) and the associated flow rule EVP models based on the MCC framework (see also Kutter and Sathialingam [27]) are incapable of capturing the 'narrow region'. To predict such a behavior, incorporation of the additional model parameters in the MCC framework are also frequent (see also Liu and Carter [42]). However, in this paper, we obtained such a 'narrow region' of the natural clay without any extra model parameters.

*4.2. San Francisco Bay Mud Clay*

From Lacerda [39]; Kaliakin and Dafalias [26], we obtain EVP model parameters for the San Francisco Bay Mud (SFBM) natural clay. In this section, we compare the EVP models' predictions with the experimentally observed undrained triaxial test results for the relaxation test. Additionally, we demonstrate herein SR-I-5 test data (see also Lacerda [39]). The water content, the liquid limit, the plastic limit, and the plasticity index of this clay sample are 88–93%, 88.4–90%, 35–44% and 45–55 %, respectively. Moreover, the specific gravity of the SFBM clay is 2.66–2.75. Also, the isotropic consolidation pressure of the clay sample is 78.4. Additionally, the initial void ratio and its equivalent mean pressure are 1.30 and 156.9 kPa, respectively (see also Kaliakin and Dafalias [26]). In Table 3, we present the shear and the relaxation phase, axial strain, strain rate and the duration of the test.

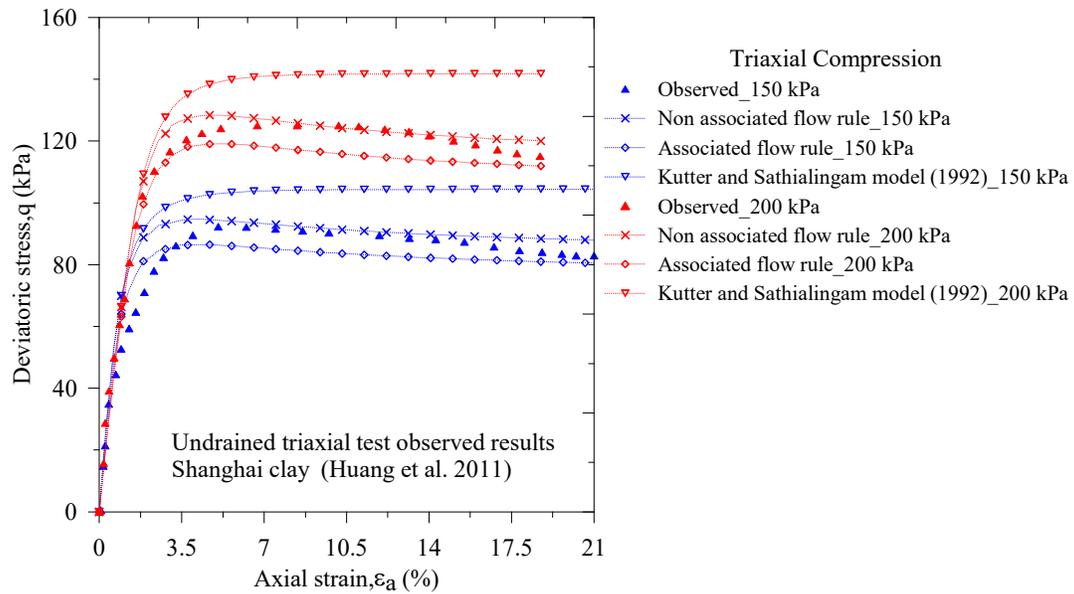

(**a**)

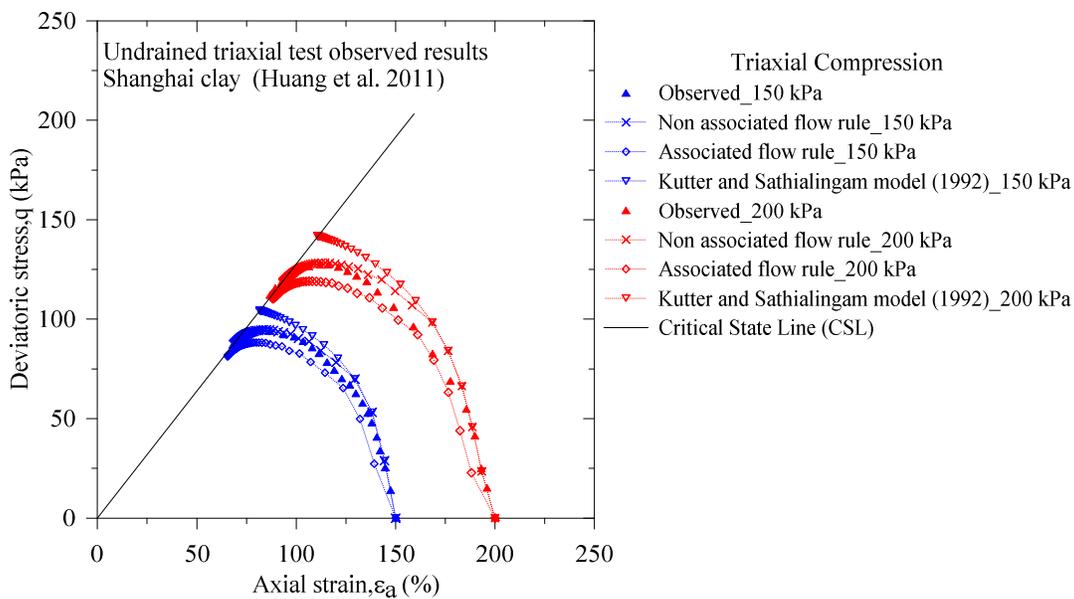

(**b**)

**Figure 9.** Comparison of observed and predicted triaxial compression test results of the Shanghai Clay (**a**) deviatoric stress-axial strain and (**b**) stress path.

In Figure 10, we present a comparison of the measured data, and EVP models predicted responses. For the EVP models, we also compare the flow rule effect considering the associated flow rule (AFR) and the non-associated flow rule (NAFR). We observe that the NAFR based EVP model capture well the experimental response. It is worth mentioning that Islam et al. [14] also find an identical flow rule effects for the extended Modified Cam Clay equivalent surface.

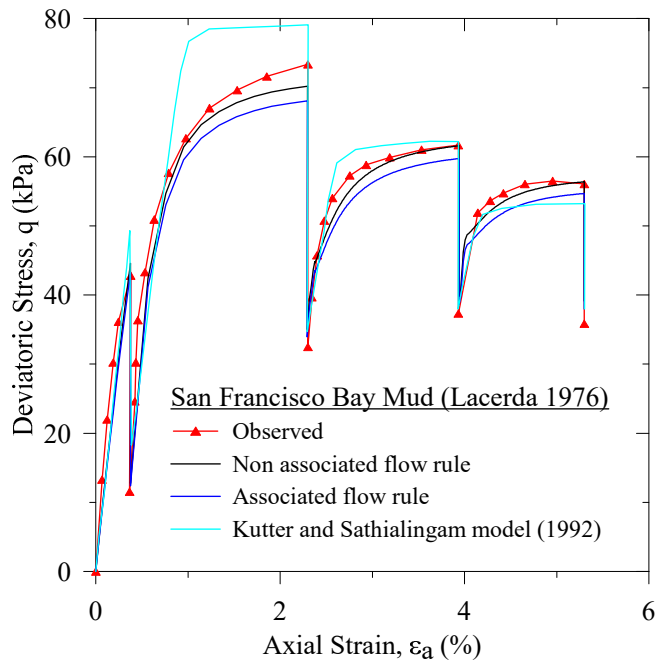

(**a**)

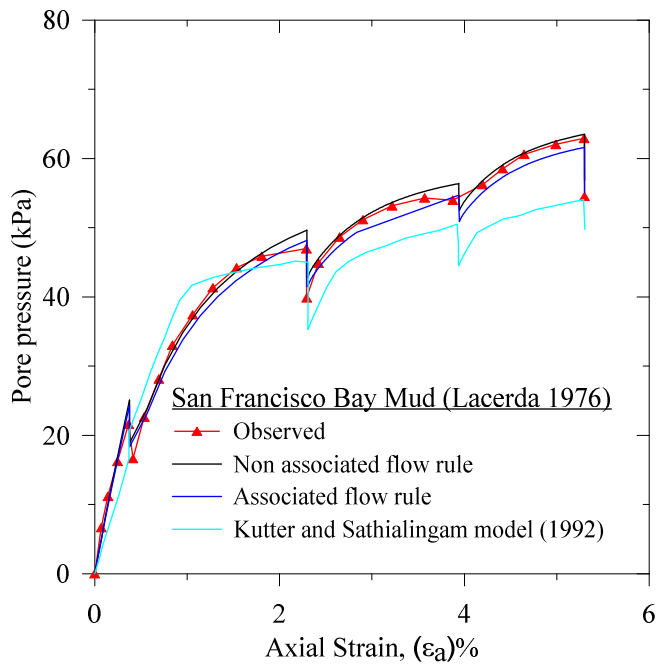

(**b**)

**Figure 10.** Comparison of observed and predicted stress relaxation triaxial test results of the San Figure 14. also find identical flow rule effects for the extended Modified Cam Clay equivalent surface.

Table 3. Stress relaxation test on the San Francisco Bay Mud clay.

| Phase | 1 | 2 | 3 | 4 | 5 | 6 | 7 | 8 |
|---|---|---|---|---|---|---|---|---|
| Test | Shear | *Relax. | Shear | *Relax. | Shear | *Relax. | Shear | *Relax. |
| $\dot{\varepsilon}_a$ | 1.5 | 0 | 1.5 | 0 | 0.0162 | 0 | 0.00081 | 0 |
| $\varepsilon_a$(%) | 0–0.38 | 0.38 | 0.38–2.3 | 2.3 | 2.3–3.94 | 3.94 | 3.94–5.3 | 5.3 |
| Time | 0.25 | 3070 | 1.28 | 1320 | 101.24 | 2700 | 1679 | 8370 |

Note: Time = minutes, *Relax. = Relaxation, $\dot{\varepsilon}_a = (\%/min)$

In Figure 11, we present a comparison of the EVP models' predicted response with the observed results of the drained triaxial compression test on the San Francisco Bay Mud clay. The triaxial sample was isotropically consolidated to the confining pressure 156.9 kPa. Additionally, the axial strain rate was 0.0031%/minute. From Figure 11, we observe that the non-associated flow rule EVP model well captured the experimentally observed results.

*4.3. Kaolin Clay*

In this section, we present the over consolidation ratio effect of the Kaolin clay (see also Herrmann et al. [40]) for the undrained triaxial compression and extension tests. It is a reconstituted clay and comprised of Snow-Cal 50 Kaolin and 5% bentonite mixture. The specific gravity of this clay is 2.64. The liquid limit and the plasticity index of this clay are 47.0% and 20.0%, respectively (see also Herrmann et al. [40]).

The initial isotropic consolidation pressure of the triaxial Kaolin clay sample was 392.2, and corresponding the initial void ratio for the unit over consolidation ratio (OCR) was 0.613 (see also Islam and Gnanendran [23]; Islam et al. [14]). We calculate the confining pressure for different OCR'S considering the initial state condition for both the triaxial compression and extension tests.

From Figure 12 and for OCR = 1, we observe that the non-associated flow rule-based EVP model prediction of the stress-strain response is satisfactory before attaining the peak. Then, after 14.0% axial strain, the under-prediction in the non-associated flow rule-based EVP model is 1.05%. Additionally, such a magnitude comparing with experimental results for the associated flow rule are 3.5% and 1.40%. We also observe a similar prediction for the undrained triaxial compression with OCR = 2, 4 and 6 for the stress-strain responses. For the triaxial extension test, we present a similar comparison for OCR = 1 and 2. From such a contrast of EVP models with the measured experimental results illustrate the effect of flow rule in the EVP models predictions.

Additionally, for OCR = 1 and 2, in Figures 12 b and 12 c, we present the pore pressure response and the stress path response, respectively considering the triaxial compression and the triaxial extension test. We also find that the non-associated flow rule prediction is close to the experimental responses.

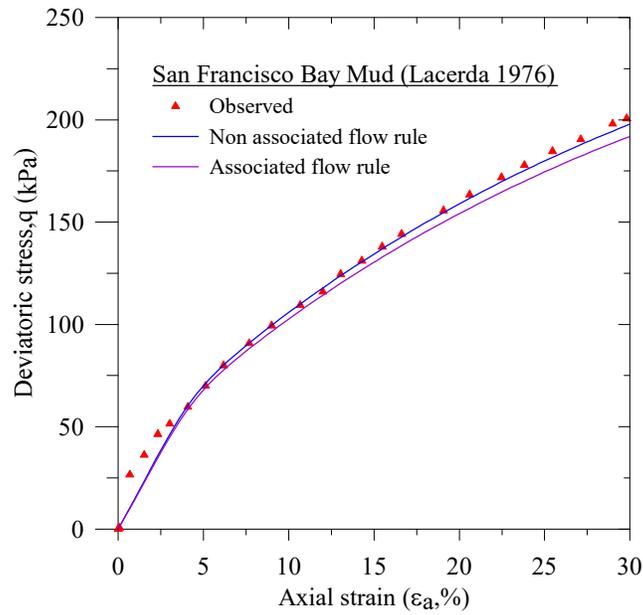

(**a**)

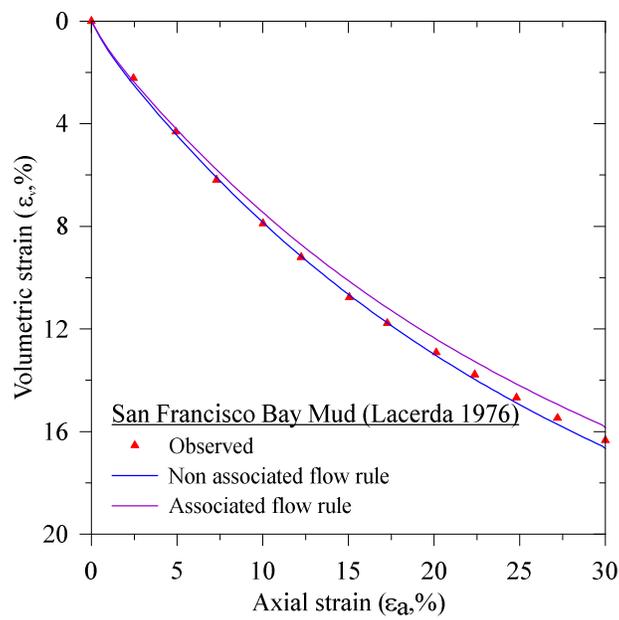

(**b**)

**Figure 11.** Comparison of the measured and the predicted consolidated drained triaxial compression tests on the San Francisco Bay Mud clay: (**a**) the deviatoric stress vs. the axial strain and (**b**) the volumetric strain vs. the axial strain.

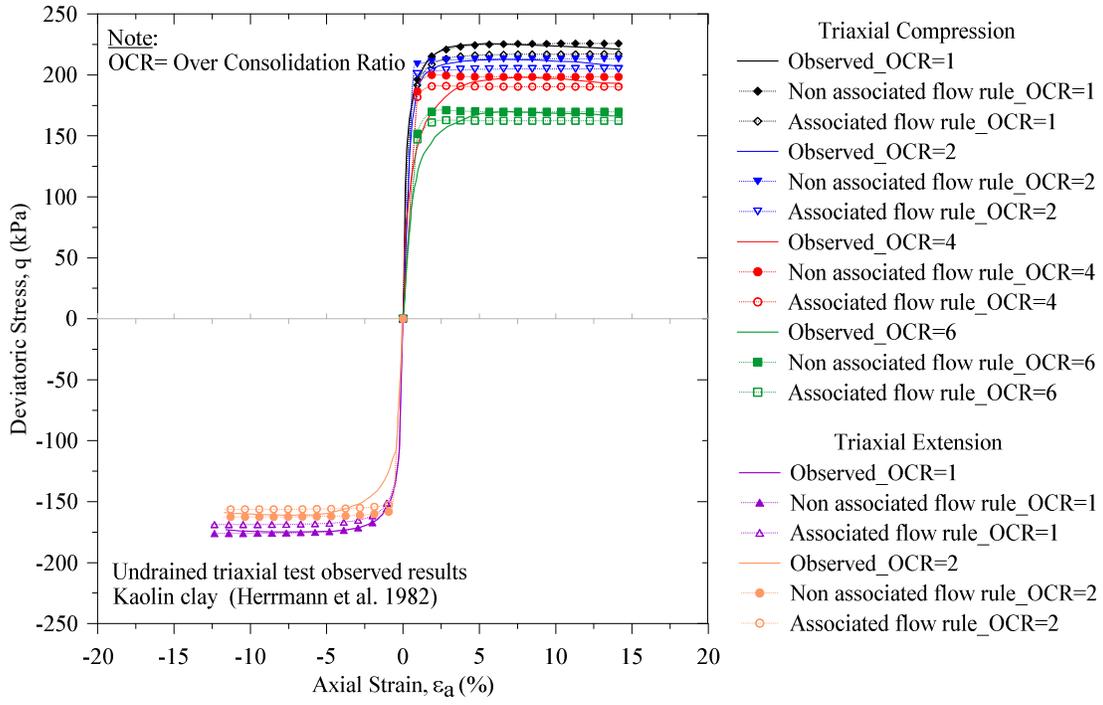

(**a**)

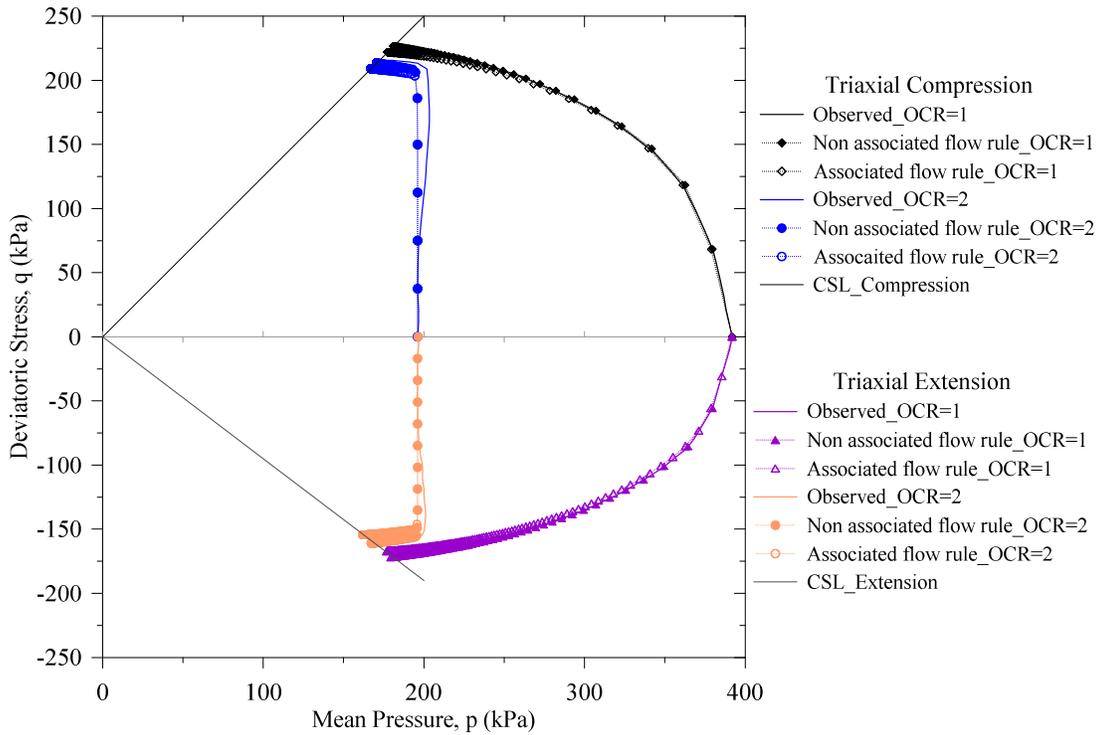

(**b**)

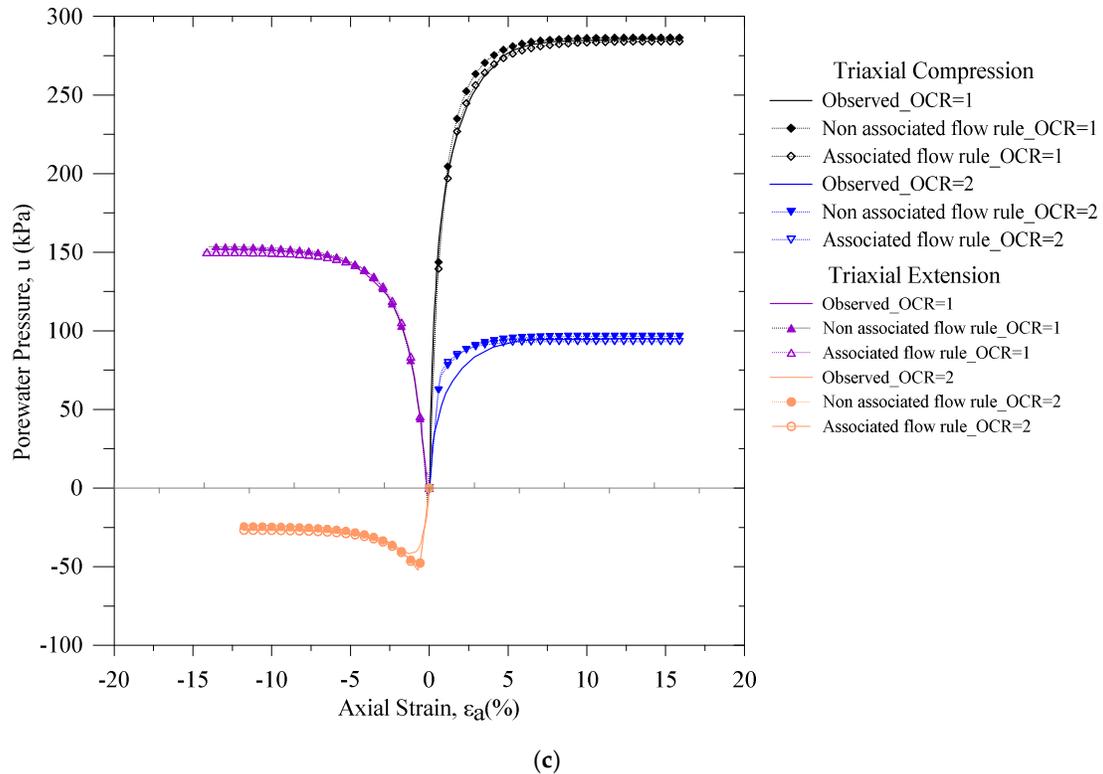

(**c**)

**Figure 12.** Comparison of observed and predicted triaxial compression test results of Kaolin Clay (**a**) Deviatoric stress-axial strain and (**b**) stress path and (**c**) pore pressure.

## 5. Application of the EVP Models

We use the EVP models in this paper to capture the long-term behavior of the Nerang-Broadbeach Roadway (NBR) embankment in Australia. Islam et al. [20] presented details of geotechnical properties, subsurface and geology, the geometry of the NBR embankment sections, measured instrumentation data for the settlement plates and piezometers, methodologies to determine the model parameters and laboratory, as well as field measured values. In this paper, we compare 590 days measured data of a surcharged-preloading section with the EVP models' predicted response considering the associated and the non-associated flow rule to illustrate the flow rule effect. Additionally, we also compare the Modified Cam Clay (MCC) predicted results with the EVP models' response to demonstrate the viscous response of the clay.

We present the geometry of the finite element section for the NBR embankment in Figure 13. Additionally, to reduce the boundary effect in numerical simulations, we extend the width and depth of the embankment. Moreover, to avoid the instabilities in the finite element simulations, we implement surcharge preloading incrementally. Furthermore, 3.0 m preloading was applied in 370 days. Then, 1.0 m additional surcharged load added and monitored for 220 days. We demonstrate the model parameter in Table 4. Moreover, in Figure 14, we present a comparison of the observed and FE predicted responses. We observe that the non-associated flow rule model well captured the measured settlement plate response compared to the associated flow rule model. Additionally, from the comparison of the MCC model with the EVP models, it is evident that the MCC model underpredicts the long-term behavior of the embankment. Such underprediction in the MCC model developed due to the exclusion of the viscous behavior of clays. Additionally, we also observe that for the long-term monitoring of EVP models, the flow rule effect is significant. We also noticed a similar response in the *Results and Discussion* Section for natural clays.

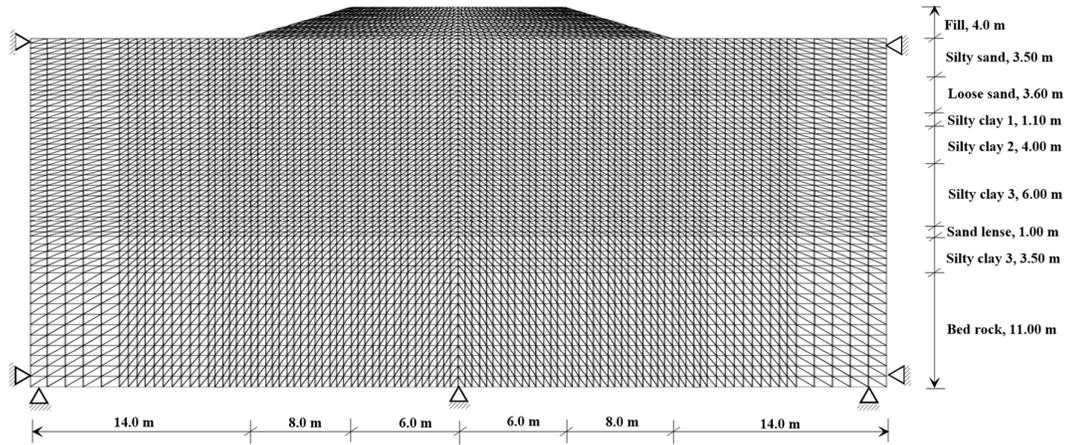

**Figure 13.** Finite element geometry of the Nerang Broadbeach Roadway embankment.

**Table 4.** Model parameters for the Nerang Broadbeach Roadway embankment.

| Soil Layer | $M$ | $\lambda$ | $\kappa$ | $e_N$ | $p_c$ (kPa) | $C_\alpha$ | $R$ |
|---|---|---|---|---|---|---|---|
| Fill | \multicolumn{6}{c}{$E' = 3{,}000$ kPa, $\varphi' = 30^0$, $c' = 5.0$ kPa} | --- |
| Silty sand | \multicolumn{6}{c}{$E' = 5{,}000$ kPa, $\varphi' = 35^0$, $c' = 2.5$ kPa} | --- |
| Loose sand | \multicolumn{6}{c}{$E' = 7{,}000$ kPa, $\varphi' = 33^0$, $c' = 1.5$ kPa} | --- |
| Silty clay 1 | 1.28 | 0.36 | 0.060 | 2.10 | 159.52 | 0.029 | 2.10 |
| Silty clay 2 | 1.25 | 0.42 | 0.043 | 3.73 | 105.36 | 0.033 | 2.10 |
| Silty clay 3 | 1.20 | 0.29 | 0.030 | 2.61 | 132.20 | 0.023 | 2.10 |
| Sand lense | \multicolumn{6}{c}{$E' = 3{,}000$ kPa, $\varphi' = 35^0$, $c' = 5.0$ kPa} | --- |
| Silty clay 3 | 1.20 | 0.29 | 0.030 | 2.61 | 287.18 | 0.023 | 2.10 |
| Bedrock | \multicolumn{6}{c}{$E' = 15{,}000$ kPa, $\varphi' = 36^0$, $c' = 50.0$ kPa} | --- |

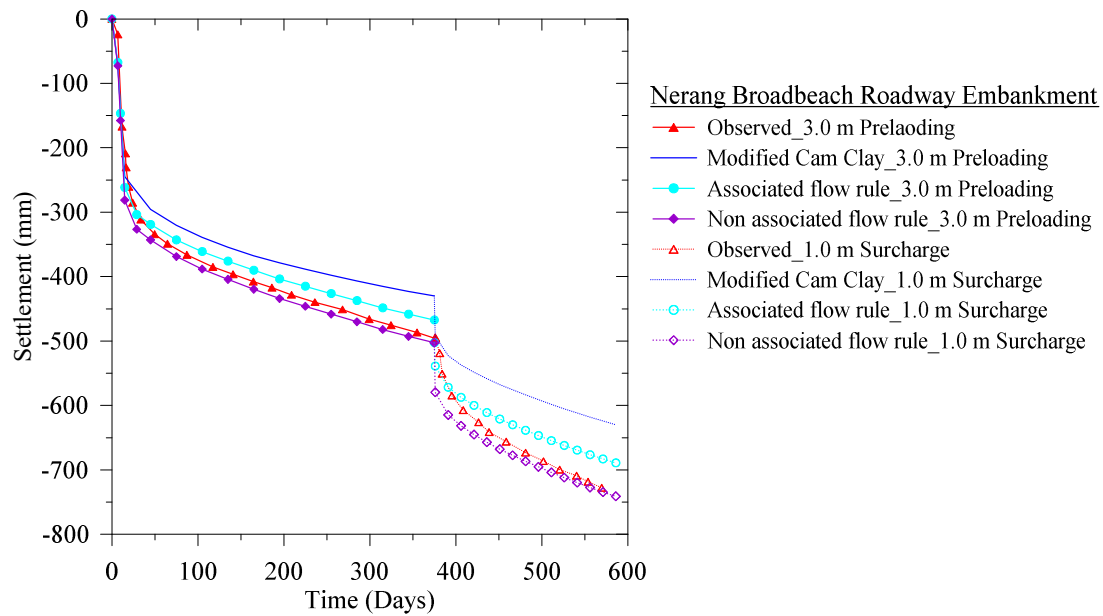

**Figure 14.** Comparison of observed and predicted settlement response of the Nerang Broadbeach Roadway embankment's surcharge-preloading.

## 6. Conclusions

In this paper, we presented a non-associated flow rule (NAFR)-based three surfaces elasto-viscoplastic model. Assuming the potential surface and the reference surface are identical, the NAFR model is also reduced to the two-surfaces associated flow rule mode. Both EVP models' surfaces comprises two ellipses, while the surface shape parameter $(R)$ controls each surface shape. Again, considering $R = 2$, we also find the extended Modified Cam Clay equivalent surface shape. Additionally, to obtain a realistic non-circular surface in the $\pi$-plane, we also included the $b$-value to define the slope of the critical state line. Thereby, by changing the $b$-value, any stress path can also be attained.

It is worth mentioning that EVP models in this paper require only seven parameters. Also, all model parameters are well known in textbooks, and all of them can be deduced from simple laboratory tests. Additionally, in the present models' formulations, we did not consider any fittings parameters. It is to note that EVP models in this paper are formulated for the isotropic clay, and any other state additional model parameters with associated modification are essential. Moreover, EVP models herein are not capable of capturing the structured behavior of highly sensitive clay, which require further changes with extra model parameters. Also, the present model may not be the best fit for the dynamic performance of clay. Furthermore, the EVP model herein is limited to the Darcy type and immiscible fluid flow.

For validation of EVP models, we considered a wide range of triaxial tests for natural and reconstituted clays. From comparisons of observed and predicted responses, we found that the non-associated flow rule EVP model well captured experimental results. Also, we found that the flow rule effect is noticeable for the natural clays than the reconstituted clay. Additionally, we implemented EVP models in a coupled consolidated finite element solver. Then, we used EVP models and the Modified Cam Clay (MCC) model to predict long-term monitoring data of the Nerang Broadbeach Roadway embankment. We also found that the non-associated flow rule EVP model well captured the settlement plate measured response compared to the associated flow rule EVP model and the MCC model.

**Author Contributions:** Conceptualization, M.N.I & C.T.G.; Investigation, M.N.I.; Methodology, M.N.I.; Supervision, C.T.G.; Validation, M.N.I.; Writing – original draft, M.N.I.; Writing – review & editing, M.N.I & C.T.G.

**Funding:** This research received no external funding.

**Acknowledgments:** The first author was financially supported during his research at the University of New South Wales, Canberra, Australia.

**Conflicts of Interest:** The authors declare no conflict of interest.

**Appendix A: Derivation of $\Phi$**

Islam and Gnanendran [23] presented a derivation of $\Phi$ for the associated flow rule-based EVP model. In this paper, we demonstrated the derivation of $\Phi$ for the non-associated flow rule considering the composite surface obtaining from the undrained triaxial compression test. Additionally, Islam et al. [14] also illustrated $\Phi$ for the Modified Cam Clay equivalent single surface model. Also, we discussed the limitations of the single surface over the composite surface base model in Section *Bounding surfaces of the EVP model.*

Considering the consolidation test of clay, we present the void ratio and the natural logarithm of mean pressure space relation (see Schofield and Wroth [21]) in Figure A1. Additionally, we obtain the change of void ratio with respect to time (after Schofield and Wroth [21] and considering viscosity (Islam and Gnanendran [23])

$$\frac{de}{dt} = -\frac{\alpha}{\bar{t}} exp\left(\frac{e - \bar{e}}{\alpha}\right). \tag{A1}$$

In Equation (A1), $\bar{t}$ represents the arbitrary time, which is not a model parameter. Following Borja and Kavazanjian [17], we introduced $\bar{t}$ in the model formulation (see also Kutter and Sathialingam [27]). It is worth mentioning that in our model formulation, the $\lambda$-line (see also Figure A1) at the reference time represents the initial bounding surface (IBS) while due to creep at time increment ($\Delta t$), the $\lambda$-line will gyrate the IBS and a new bounding surface will generate.

Again, considering the viscoplasticity, we obtain the volumetric component of the viscoplastic strain rate $\left(\dot{\varepsilon}_v^{vp}\right)$ as:

$$\dot{\varepsilon}_v^{vp} = -\frac{de}{dt}\frac{1}{1+e_0}. \tag{A2}$$

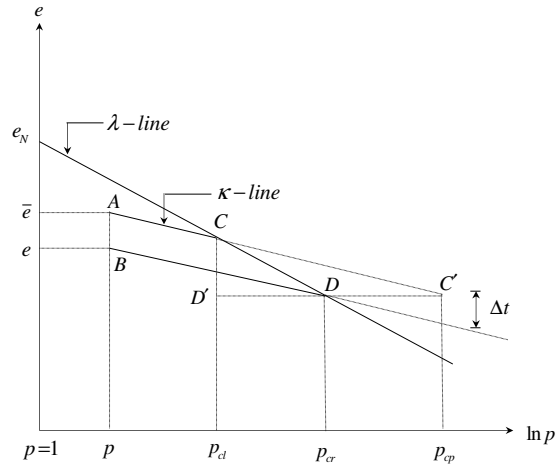

**Figure A1.** Undrained triaxial test path for the virgin consolidated clay for the normal consolidation line and the swelling consolidation line representing the initial surfaces.

From Figure A1, we also find (see also Roscoe and Burland [10]):

$$\bar{e} - e = (\lambda - \kappa)ln\left(\frac{p_{cl}}{p_{cr}}\right). \tag{A3}$$

Now, combining Equations (A1) to (A3), we find:

$$\dot{\varepsilon}_v^{vp} = \frac{\alpha}{\bar{t}(1+e_0)}\left(\frac{p_{cl}}{p_{cr}}\right)^{\frac{\lambda-\kappa}{\alpha}}. \tag{A4}$$

In the triaxial stress space, $\dot{\varepsilon}_v^{vp}$ also can be written as:

$$\dot{\varepsilon}_v^{vp} = \Phi\frac{\partial f_p}{\partial p_p}. \tag{A5}$$

By comparing Equations (A4) and (A5), then separating $\Phi$ for calculation of the viscoplastic strain component results in the flow rule independent elasto-viscoplastic model, which violates the flow rule theory (see also Islam and Gnanendran [23]). Therefore, for the non-associated flow rule, to obtain $\Phi$, we find $\frac{\partial f_p}{\partial p_p}$ considering the one-dimensional compression test criterion for relevant flow rule as follows:

$$\Phi = \frac{\alpha_0}{\bar{t}(1+e_0)}\left(\frac{p_{cl}}{p_{cr}}\right)^{\frac{\lambda-\kappa}{\alpha}}\frac{1}{2p_{cp}\left[\frac{1}{\varsigma}-\frac{1}{R}\right]}, \tag{A6}$$

$$\varsigma = \frac{p_{cp}}{p_p} = \frac{-1+(R-1)\sqrt{1+R(R-2)\left(\frac{\eta_0}{M}\right)^2}}{R-2}, \tag{A7}$$

$$\eta_0 = \frac{\left(6(R-1)^2(\lambda-\kappa)-2\sqrt{9(\lambda-\kappa)^2(R-1)^2+(2\lambda M)^2}\right)\lambda M^2}{9(\lambda-\kappa)^2(R^4-4R^3+5R^2-2R)-(2\lambda M)^2}. \tag{A8}$$

Again, from Figure A1, we also find expression for $p_{cr}$ and $p_{cp}$ (see also Islam et al. [14]):

$$p_{cr} = exp\left(\frac{e_N - e - \kappa\,lnp}{\lambda-\kappa}\right), \tag{A9}$$

$$p_{cp} = \frac{p_{cr}^{\frac{\lambda}{\kappa}}}{p_{cl}^{\frac{\lambda-\kappa}{\kappa}}}. \tag{A10}$$

## Appendix B: Derivation of $\dot{\varepsilon}_{ij}^{vp}$

In Equation (24), we defined $\dot{\varepsilon}_{ij}^{vp} = \langle \Phi(F) \rangle \frac{\partial f_p}{\partial \sigma'_{ij}}$ while in Appendix A presented expression of $\langle \Phi(F) \rangle$. Again, using the chain rule, we find $\frac{\partial f_p}{\partial \sigma'_{ij}}$ as follows:

$$\frac{\partial f_p}{\partial \sigma'_{ij}} = \frac{\partial f_p}{\partial p_p}\frac{\partial p_p}{\partial \sigma'_{ij}} + \frac{\partial f_p}{\partial q_p}\frac{\partial q_p}{\partial \sigma'_{ij}} + \frac{\partial f_p}{\partial M}\frac{\partial M}{\partial b}\frac{\partial b}{\partial \sigma'_{ij}}, \tag{A11}$$

$$\frac{\partial f_p}{\partial p_p} = 2\left(p_p - \frac{p_{cp}}{R}\right), \tag{A12}$$

$$\frac{\partial f_p}{\partial q_p} = 2(R-1)^2 \frac{q_p}{M^2}. \tag{A13}$$

In Islam and Gnanendran [23]; Islam et al. [14] we presented $\frac{\partial p_p}{\partial \sigma'_{ij}}$; $\frac{\partial q_p}{\partial \sigma'_{ij}}$; $\frac{\partial f_p}{\partial M}$; $\frac{\partial f_p}{\partial M}$; $\frac{\partial M}{\partial b}$; $\frac{\partial b}{\partial \sigma'_{ij}}$ for the associated flow rule and assuming the single surface model. For the sake of completeness, we illustrate them for the non-associated flow rule and composite surface based EVP model as follows:

$$\frac{\partial p_p}{\partial \sigma'_{ij}} = \frac{1}{3}\delta_{ij}, \text{ where } \delta_{ij} = \begin{cases} 1 & \text{if } i = j \\ 0 & \text{if } i \neq j \end{cases}, \tag{A14}$$

$$\frac{\partial q_p}{\partial \sigma'_{ij}} = \begin{cases} \frac{3}{2q_p}\left(\acute{\sigma}_{ij}^p - p_p\delta_{ij}\right) & \text{if } i = j \\ \frac{3}{2q_p}\left(2\acute{\sigma}_{ij}^p\right) & \text{if } i \neq j \end{cases}, \tag{A15}$$

$$\frac{\partial f_p}{\partial M} = \frac{-(R-1)^2 2 q_p^2}{M^3} \quad \text{for ellipse 1}, \tag{A16}$$

$$\frac{\partial f_p}{\partial M} = \frac{-2q_p^2}{M^3} \quad \text{for ellipse 2}, \tag{A17}$$

$$\frac{\partial M}{\partial b} = \frac{3\sin\phi(2b-1)}{(\sqrt{b^2 - b + 1})[3 + (2b-1)\sin\phi]} - \frac{12\sin^2\phi(\sqrt{b^2 - b + 1})}{[3 + (2b-1)\sin\phi]^2}, \tag{A18}$$

$$\frac{\partial b}{\partial \acute{\sigma}_{11}^p} = -\frac{\acute{\sigma}_{22}^p - \acute{\sigma}_{33}^p}{\left(\acute{\sigma}_{11}^p - \acute{\sigma}_{33}^p\right)^2}, \tag{A18}$$

$$\frac{\partial b}{\partial \acute{\sigma}_{22}^p} = \frac{1}{\acute{\sigma}_{11}^p - \acute{\sigma}_{33}^p}, \tag{A19}$$

$$\frac{\partial b}{\partial \acute{\sigma}_{33}^p} = -\frac{1}{\acute{\sigma}_{11}^p - \acute{\sigma}_{33}^p} + \frac{\acute{\sigma}_{22}^p - \acute{\sigma}_{33}^p}{\left(\acute{\sigma}_{11}^p - \acute{\sigma}_{33}^p\right)^2}. \tag{A20}$$

To resolve the limitations of the strain-softening phenomena in the undrained triaxial test, among others, Liu and Carter [42] introduced a multiplication factor $\left(1 + \frac{\eta}{M+\eta}\right)$. The purpose of this fuction is to account for both the current stress ratio and the history. In this paper, considering the softening behavior, such a multiplication factor can also be implemented to calculate the viscoplastic strain rate.